\documentclass[prb,twocolumn,amsmath,amssymb,10pt,aps,longbibliography,superscriptaddress,citeautoscript,bibnotes,footinbib]{revtex4-1}

\usepackage{feynmp}
\usepackage{amsmath}
\usepackage{graphicx}
\usepackage{multirow}
\usepackage{latexsym}
\usepackage{textcomp}
\usepackage{verbatim}
\usepackage{color}
\usepackage{bm}
\usepackage{subfigure}
\usepackage{everysel}
\usepackage{keyval}
\usepackage{ragged2e}
\usepackage{dsfont}
\usepackage{amssymb}
\usepackage{enumitem}
\usepackage{pstricks}
\usepackage{mathtools}
\usepackage{braket}
\usepackage{slashed} 
\usepackage[colorlinks=true]{hyperref}

\usepackage{graphicx}
\usepackage{xcolor}
\usepackage{amsmath}
\usepackage{amsfonts}
\usepackage{bbm}

\newcommand{\osum}{{%
    \setbox0\hbox{\circ}%
    \rlap{\hbox to \wd0{\hss\sum\hss}}\box0
}}

\begin{document}

\title{Many-Body Order Parameters for Multipoles in Solids}

\author{Byungmin Kang}
\affiliation{School of Physics, Korea Institute for Advanced Study, Seoul 02455, Korea}

\author{Ken Shiozaki}
\affiliation{Yukawa Institute for Theoretical Physics, Kyoto University, Kyoto 606-8502, Japan}

\author{Gil Young Cho}
\thanks{Electronic Address: gilyoungcho@postech.ac.kr}
\affiliation{Department of Physics, Pohang University of Science and Technology (POSTECH), Pohang 37673, Republic of Korea}

\date{\today}
 
\begin{abstract} 
We propose many-body order parameters for bulk multipoles in crystalline systems, which originate from the non-trivial topology of the nested Wilson loop spectrum. The many-body order parameters are designed to measure multipolar charge distribution in a crystalline unit cell, and they match the localized corner charge originating from the multipoles. We provide analytic arguments and numerical proof for the order parameters. Furthermore, we show that the many-body order parameters faithfully measure the physical multipole moments even when the symmetries quantizing multipoles are lost and thus the nested Wilson loop spectrum does not exactly reproduce the physical multipole moments. 
\end{abstract}

\maketitle

Theory of macroscopic polarization in a crystal lies at the heart of the modern development of topological band insulators~\cite{RevModPhys.66.899}. The theory provides a classic example of how the non-trivial topology of bulk ground states can determine the quintessential properties of the quantum phases such as symmetry-enforced boundary states~\cite{RevModPhys.82.3045, RevModPhys.83.1057}. Furthermore, it nicely illustrates how the associated topology can and cannot be diagnosed; the topology cannot be measured by a local operator, but only through a gauge-invariant number defined in a momentum space referred as the Zak phase~\cite{PhysRevLett.62.2747}, or through a non-local many-body order parameter~\cite{PhysRevLett.80.1800}. In particular, the discovery of the many-body order parameters, or equivalently ``invariants", for various topological insulators drastically increased our understanding of topological states and expanded the range of the search for the topological states beyond free-fermion limits, e.g., strongly-correlated systems and bosonic models~\cite{PhysRevLett.82.370, PhysRevB.88.155121, Spin,PhysRevB.91.195142, PhysRevLett.118.216402}.   

Recently, a new class of topological states, namely higher-order topological insulators, has been discovered~\cite{Benalcazar61, PhysRevB.96.245115, HOTI_1, HOTI_2, HOTI_3, HOTI_4,  HOTI_12, HOTI_15, PhysRevB.98.201114}. Interestingly, these higher-order states are constructed out of the electric multipoles, which materialize as topologically-protected modes at the corners and hinges of the physical boundary. To characterize such higher-order topology, the so-called ``nested Wilson loop" approach~\cite{PhysRevB.96.245115} has been put forward and used to detect the quantized multipoles. However, the nested Wilson loop indices successfully work only when there exist symmetries which quantize the multipole moments~\cite{PhysRevB.96.245115} and currently there is no generic bulk measure for the multipoles working even in the absence of the symmetries. Also, a new class of higher-order topological insulator has been introduced under the name an anomalous topological insulator\cite{PhysRevB.98.201114} whose non-trivial multipole moments cannot be diagnosed by a naive application of the nested Wilson loop approach. Moreover, the nested Wilson loop is genuinely a free-fermion quantity and thus cannot be a generic diagnosis for the higher-order topology in a general many-body condensed matter system. This opens a question how to measure multipole moments going beyond the nested Wilson loop picture.  

Motivated from this, we propose a set of many-body order parameters for the multipoles in crystalline systems, which succsefully measure the multipole moments even when the quantizing symmetries are lost so that the nested Wilson loop indices cannot be used to detect the true bulk moments. In stark contrast to the nested Wilson loop approaches, we utilize real-space charge distribution, which can be evaluated even for strongly-correlated systems. To this end, we extend Resta's general definition of polarization~\cite{PhysRevLett.80.1800} to the multipoles:  
\begin{align}
&P_{x} = \frac{1}{2\pi} \text{Im}\left[ \log \langle \hat{U}_1 \rangle \right], ~~ \hat{U}_{1} = e^{2\pi i \sum_{x} \hat{p}_{x}}, 
\label{Analytic0}
\end{align}
where $\hat{p}_x = \frac{x \hat{n}(x)}{L_x}$ is the polarization density (relative to $x=0$) in the system of the length $L_x$ with $\hat{n}(x)$ being the electron number at site $x$, and the sum is running over the whole system. Here the expectation value $\langle \hat{U}_1 \rangle$ is over the many-body ground state subject to the periodic boundary condition, i.e., $x \sim x+L_x$. Compared to the Berry-phase expression~\cite{PhysRevLett.62.2747} defined in momentum space, the above formula Eq.~\eqref{Analytic0} provides a clearer picture of the macroscopic polarization $P_x$ by relating it to the local microscopic polarization density $\hat{p}_x$. It also has played an important role in developing theories of related topological states\cite{Oshikawa-Senthil, lu2017filling} and other strongly-correlated systems, which is known as generalized Hastings-Oshikawa-Lieb-Schultz-Mattis theorem\cite{Oshikawa,Affleck1986,parameswaran2013topological}. 

Here, we define the macroscopic quadrupole moment in a crystal as following: 
\begin{align}
&Q_{xy} = \frac{1}{2\pi} \text{Im}\left[ \log \langle \hat{U}_2 \rangle \right], ~~ \hat{U}_{2} = e^{2\pi i \sum_{\bm{r}} \hat{q}_{xy}(\bm{r})}.
\label{Analytic1}
\end{align}
Here $\hat{q}_{xy} = \frac{xy}{L_x L_y} \hat{n}(\bm{r})$ is the quadrupole moment density (relative to $x=y=0$) per unit cell at the site $\bm{r}$. Similar to Eq.~\eqref{Analytic0}, $L_x$ and $L_y$ are the sizes of the system, and the sum is over $(x,y) \in (0, L_x] \times (0, L_y]$. Analogously, the macroscopic octupole moment is defined as: 
\begin{align}
&O_{xyz} = \frac{1}{2\pi} \text{Im}\left[ \log \langle \hat{U}_3 \rangle \right], ~~ \hat{U}_{3} = e^{2\pi i \sum_{\bm{r}} \hat{o}_{xyz}(\bm{r})}. 
\label{Analytic2}
\end{align}
Here $\hat{o}_{xyz} = \frac{xyz}{L_x L_y L_z} \hat{n}(\bm{r})$ is the octupole moment density per unit cell; the generalizations to higher-order poles follows immediately. As in Resta's formula, we use a many-body ground state defined on the torus with periodic boundary conditions and each multipole moment density is summed over the whole space. Our real-space formula provides a complementary view to the momentum space approach, the nested Wilson loop indices~\cite{PhysRevB.96.245115}. In this paper, we prove that these many-body order parameters correctly measure the multipole moments which originates from the topology of the nested Wilson loop spectrum.~\cite{PhysRevB.96.245115} 


A few remarks are in order. First, the multipole moments defined as above are ambiguous up to ``mod 1" because of the periodicity of complex phases; this ambiguity reproduces the observation made in the previous studies~\cite{Benalcazar61, PhysRevB.96.245115}. Second, we emphasize that the above expressions are truly many-body order quantities because, when the exponetials are expanded, they require to measure $\sim \sum_{\bm{r}_1, \bm{r}_2, \cdots, \bm{r}_J}\langle \hat{n}(\bm{r}_1) \hat{n}(\bm{r}_2) \cdots \hat{n}(\bm{r}_{J}) \rangle$, and this does not reduce to the product of the few-particle observables. To evaluate Eq.~\eqref{Analytic1} and Eq.~\eqref{Analytic2}, we need the full knowledge of the many-body ground state.  On the other hand, the expressions are gauge-invariant, i.e., it is independent of the basis choice, and thus they measure physical quantities.  

We can also show that the expressions Eq.~\eqref{Analytic1} and Eq.~\eqref{Analytic2} are invariant under the superposition of the trivial atomic insulator to the system. For example, we take Eq.~\eqref{Analytic1} and $L_x \cdot L_y$ being odd. Let us first consider a ground state $|GS_{\text{triv}}\rangle = \otimes_x |\hat{n}_x = 1 \rangle$ of an atomic trivial insulator, which has $\langle \hat{U}_2 \rangle = 1$. When such a ground state is superposed with the topological state $|GS _{\text{top}}\rangle$, i.e., $|GS_{\text{new}}\rangle = |GS_{\text{top}}\rangle \otimes |GS_{\text{triv}}\rangle$, the quadrupole moment remains the same as before
\begin{align}
\text{Im} \log \langle GS_{\text{top}} |\hat{U}_2 |GS_{\text{top}} \rangle = \text{Im} \log \langle GS_{\text{new}} |\hat{U}_2 |GS_{\text{new}} \rangle, \nonumber 
\end{align}
and hence showing the stability against addition of trivial bands (without turning on the coupling to the topological states).~\cite{comment1}  This shows that our invariants Eq.~\eqref{Analytic1} and Eq.~\eqref{Analytic2} measure certain topological properties of the insulators. To further identify them as the ``measure" for the multipoles, we need a few more steps.  

\textbf{1. Analytic Arguments}: Now we argue that Eq.~\eqref{Analytic1} and Eq.~\eqref{Analytic2} \textit{are} the physical multipole moments. We will concentrate on the quadrupole moment Eq.~\eqref{Analytic1} for clarity. Below we choose our system to be on a torus $(x,y) \in (0, L_x] \times (0, L_y]$ and subject to the periodic boundary conditions $x+ L_x \sim x$ and $y+ L_y \sim y$. The effect of other coordinate parametrization is discussed in the supplemental material.\cite{suppl_ref}

First, we note that the $Q_{xy}$ defined as above transforms under the crystalline symmetries as the physical quadrupole moment. For example, under the mirrors $\{\mathcal{M}_x, \mathcal{M}_y\}$, we see that $\hat{U}_2 \to \hat{U}_2^*$, where asterisk denotes the complex conjugation, and thus $Q_{xy} \to -Q_{xy}$. (We defer a more careful analysis of the symmetry actions including $C_4$ symmetry to the supplemental material\cite{suppl_ref}.)


Second, we can also intuitively understand why the polarization must vanish for the quadrupole moment to be well-defined. To this end, we perform $x \to x+ L_x$ globally, which maps $\hat{U}_2 \to \hat{U}_2' = \hat{U}_2 \hat{U}_{1;y} $
where $\hat{U}_{1;y} = \exp [ 2\pi i/L_y  \sum_{y} y\hat{n}(\bm{r}) ] = \exp (2\pi i \hat{\mathcal{P}}_y )$. Although $\langle \hat{U}_2 \rangle \neq \langle \hat{U}_2' \rangle$ in general (because the ground state is not generally an eigenstate of $\hat{U}_{1;y}$), we find 
\begin{align}
\langle \hat{U}_2' \rangle = \langle \hat{U}_2 \rangle \langle \hat{U}_{1;y} \rangle + \mathcal{O}(\frac{1}{E_{\text{gap}}}), \nonumber 
\end{align} 
where $E_{\text{gap}}$ is the excitation gap, which is inversely proportional to the correlation length~\cite{PhysRevLett.82.370, Watanabe-Oshikawa, suppl_ref}. In the ideal limit $E_{\text{gap}} \to \infty$, we can see that  the quadrupole moment, which is the imaginary part of $\langle \hat{U}_2 \rangle$, is well-defined when the polarization vanishes. Similarly, the total polarization $\hat{\mathcal{P}}_x$ along $x$ must vanish to have a well-defined quadrupole moment. By passing, we note that this invariance of $\langle \hat{U}_2 \rangle$ up to the perturbative correction is \textit{weaker} than the original Resta's formula Eq.~\eqref{Analytic0} where $x \to x+ L_x$ is an exact symmetry\cite{suppl_ref}. Though we have shown that the polarization at $E_{\text{gap}} \to \infty$ must vanish to have a well-defined quadrupole moment, the polarization is invariant under the adiabatic change of the gap when the quantizing symmetries present and thus the same conclusion should apply to the finite-gap systems. 


Finally, we can gain an important intuition from the phenomenological effective theory of the multipole moments. Using the standard field theory technique, we note  
\begin{align}
\langle GS| \hat{U}_a | GS \rangle = \text{Tr} \bigg[ \frac{e^{-\beta H}}{\mathcal{Z}_0} \hat{U}_a \bigg]\Big|_{\beta \to \infty} \propto e^{iS_{eff}[A_{0;a} (\bm{r})]},  
\label{Eff}
\end{align}
in which $a = 1, 2, 3$ depends on the particular multipole that we are interested in, $A_{0;a}(\bm{r})$ is the gauge potential configuration generated by the unitary $\hat{U}_a$~\cite{suppl_ref}, and $\mathcal{Z}_0$ is the corresponding partition function. For the dipole case, we have 
\begin{align}
A_{0;a=1} (x, \tau) = 2\pi \frac{x}{L_x} \delta(\tau), ~~S_{eff} = \int^{L_x}_0 dx\int^{\beta}_0 d\tau ~ P_{x} \cdot E_x. \nonumber   
\end{align}
Here we used the semi-classical electromagnetism response of the polarization $S_{eff}= \int \vec{P}\cdot \vec{E}$. Remarkably, despite of its semi-classical origin, we note that it is equivalent to the fully-quantum theory\cite{Jackiw, Qi_Hughes_Zhang} of one-dimensional topological insulator. Explicitly evaluating the value of the effective action, we find $\langle \hat{U}_1 \rangle \propto e^{2\pi i P_x}$. This can be further identified with the edge charge, which we derive by solving the equation of motion $q_{\text{edge}} = \int dx ~\frac{\delta \mathcal{L}_{eff}}{\delta A_0} = P_x$.

We generalize this to the quadrupole case, whose effective response is\cite{suppl_ref} 
\begin{align}
S_{eff} = \int^{L_x}_0 dx \int^{L_y}_0 dy \int^{\beta}_0 d\tau ~ \frac{Q_{xy}}{2} (\partial_y E_x + \partial_x E_y), 
\end{align}
where $Q_{xy}$ is the physical quadrupole moment density and $E_{x,y}$ are electric fields. This effective theory captures the defining characteristics of the quadrupole insulator, including the corner charge and the charge current pattern generated from an adiabatic Thouless pumping\cite{suppl_ref}. Implementing $A_{0, a=2} = \frac{2\pi}{L_x L_y} xy \delta (\tau) $, we find 
$
\langle \hat{U}_2 \rangle \propto e^{2\pi i Q_{xy}}. 
$  
It is straightforward to show that this quadrupole moment $Q_{xy}$ is precisely the corner charge $q_c$, i.e., $Q_{xy} =q_c$, by solving the equation of motion for the probe gauge field $\delta A_0$ as done in the polarization case. Similar reasoning can be applied to the octupole case, too. In summary, our many-body order parameters agree with the corner charges $q_c$, i.e., $\langle \hat{U}_a \rangle = \exp(2\pi i q_c ), a= 1,2,3$, which are the physical manifestation of the multipole moments even in the absence of quantizing symmetries. 

- \textbf{Generalizations}: With the effective theory in hand, we can now generalize our order parameters to arbitrary boundary condition, since the above relation Eq.~\eqref{Eff} holds independently of which boundary condition being used. Furthermore, we can generalize the unitary $\hat{U}_a$ to other unitaries saturating the effective action. For example, we can use
\begin{align}
\hat{V}_1 (l) & = 
\begin{cases}
\exp\Big[\frac{2\pi i}{l} \sum_{x} x \hat{n}(x)\Big] \quad \text{ for } x \in (0, l] \\ 
1 \qquad \qquad \qquad \qquad \quad \text{ for } x \in (l, L_x],
\end{cases}
\label{Analytic_G1}
\end{align}
to measure the polarization\cite{Affleck1986}. Here the whole system may be subject to the open boundary condition or to the periodic boundary condition. Similarly, we can define the generalized measure for the quadrupole moment as following: 
\begin{align}
\hat{V}_2 (l) = 
\begin{cases}
\exp\Big[\frac{2\pi i}{l^2}  \sum_{\bm{r}} xy \hat{n}(\bm{r})\Big] \quad \text{ for } \bm{r} \in (0, l] \times (0, l] \\ 
1 \qquad \qquad \qquad \qquad \quad \text{ otherwise.}
\end{cases}
\label{Analytic_G2}
\end{align}
Again, the whole system may be subject to the periodic or open boundary conditions. We will see numerically that these operators work as good as the original formula.

\begin{figure}[!t]
\includegraphics[width=1.0\columnwidth]{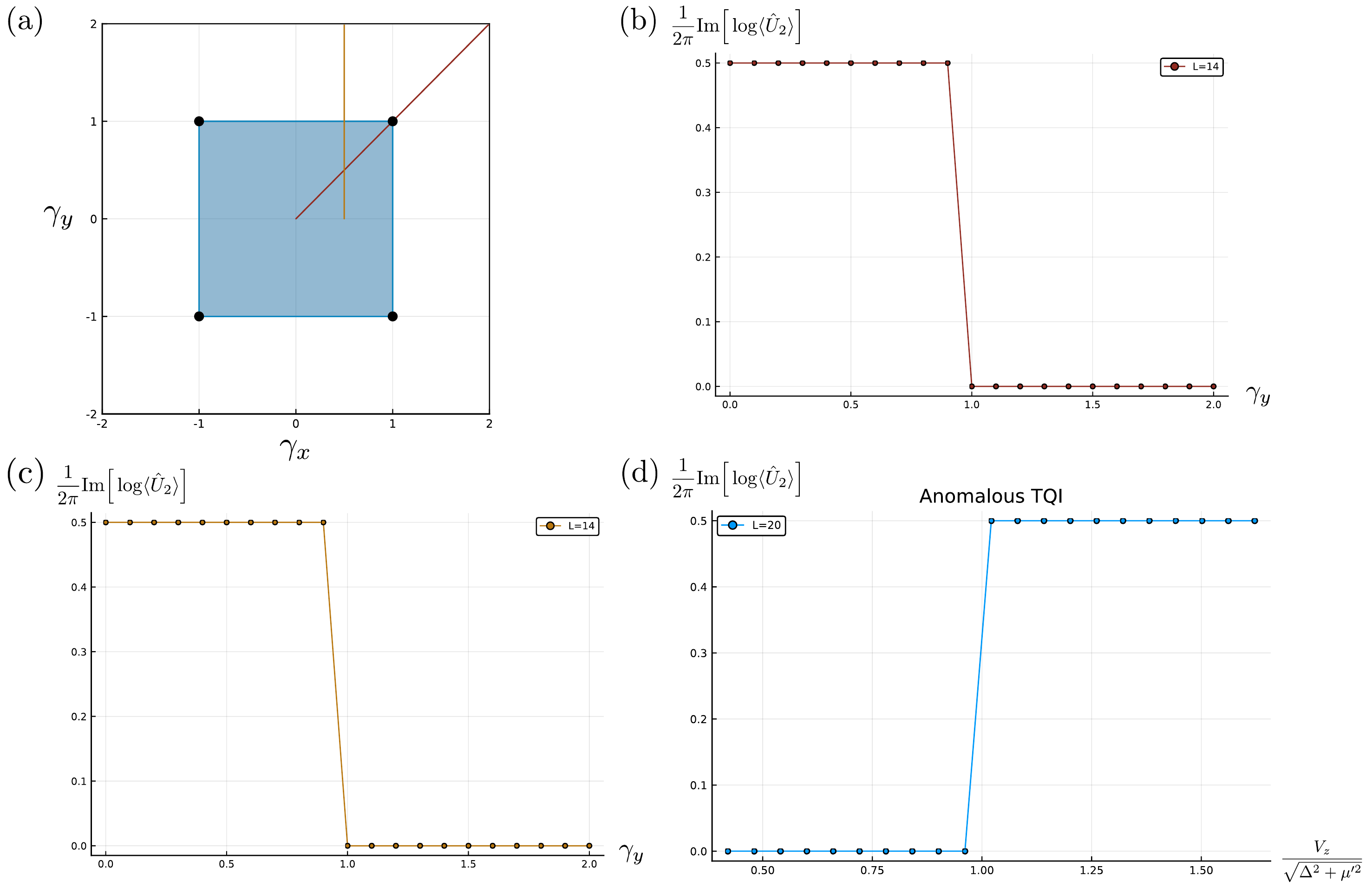}
\caption{(a) The phase diagram of the quadrupole insulator Eq.~\eqref{TQI_k_space} when $\delta=0$ and $\lambda_x = \lambda_y = 1$. The shaded region denotes topologically nontrivial phase (quadrupole moment equals 0.5) and black dots denote bulk energy gap closing points. (b) and (c) are evaluation of $Q_{xy}$ Eq.~\eqref{Analytic1} for the quadrupole insulator Eq.~\eqref{TQI_k_space}. We set $\lambda_x = \lambda_y = 1$ and $\delta = 0$, and evaluate $Q_{xy}$ (b) along the cut $\gamma_x = \gamma_y$ and (c) along the cut $\gamma_x = 0.5$. We see in both cases there is a sharp change at $\gamma_y = 1$, which is consistent with the phase diagram (a). (d) Anomalous topological quadrupole insulator.\cite{PhysRevB.98.201114} Here, the ground state is known to be topological if $\frac{V_z}{\sqrt{\Delta^2 + \mu'{}^2}}>1$  and trivial if $\frac{V_z}{\sqrt{\Delta^2 + \mu'{}^2}}<1$, and this is well captured by $\hat{U}_2$. See the supplemental material for details on the model and the numerical values of the parameters\cite{suppl_ref}.
} 

\label{U2_TQI_fig}
\end{figure}

- \textbf{Corner Charge, Nested Wilson Loop Index, and Multipole Moment}: Here we discuss a few related but distinct quantities by restricting ourselves to the quadrupolar case.  (1) corner charge $q_{\text{c}}$, which is the total electric charge near the corner when the system is subject to the open boundary conditions. (2) Wannier-sector polarizations~\cite{PhysRevB.96.245115} $\{ p_x^\omega,  p_y^\omega \}$ (when they can be defined) and associated quadrupole moment $Q^{\omega}_{xy} = 2 p_x^{\omega}p_y^{\omega}$, which are defined by nested Wilson loops, and (3) physical quadrupole moment $Q_{xy}^{\text{ph}}$. In general, all these three can be different from each other.  

However, if the bulk and boundary-alone polarizations are absent (so that the quadrupole moments are well-defined) and if the quadrupolar charge distribution per unit cell is the only source of the corner charge, then the corner charge $q_c$ would agree with a sensible quadrupole moment $Q_{xy}^{\text{ph}}$.\cite{PhysRevB.96.245115} On the other hand, the nested Wilson loop index $Q^\omega_{xy}$ does not agree with $q_{\text{c}}$ generally~\cite{PhysRevB.96.245115}. Hence, when the bulk-boundary correspondence~\cite{PhysRevB.48.4442,PhysRevB.95.035421} holds, i.e., the boundary polarization in the systems of the open boundary condition is fully determined by the bulk quadrupole moments~\cite{PhysRevB.96.245115}, we expect generically
\begin{align}
q_{\text{c}} \equiv Q^{\text{ph}}_{xy} \neq Q^\omega_{xy}, \nonumber
\end{align}
in the absence of quantizing symmetries. 

In this regard, our key claim is that, if the gap in the nested Wilson loop spectrum is present (i.e., $Q^{\omega}_{xy}$ can be defined),\cite{PhysRevB.96.245115} our many-body order parameter $Q_{xy}$ in Eq.~\eqref{Analytic1} measures the physical observable $Q^{\text{ph}}_{xy} = q_c$, i.e., $Q_{xy} = Q^{\text{ph}}_{xy}=q_c$, which may not agree with the nested Wilson loop $Q^\omega_{xy}$ \textit{in the absence of quantizing symmetries}. 

Note that, when the gap in the nested Wilson loop spectrum is absent and thus the quadrupole moments cannot be defined from the nested Wilson loop method, e.g., some $C_4$-symmetric models in Ref.~\onlinecite{Cn_insulator}, $Q_{xy}$ does not show stable values against the change in parameters nor agree with the corner charge.\cite{suppl_ref}  


\textbf{2. Numerical Proof}: We now proceed to the numerical proof for our many-body order parameters by testing the formula on the \textit{non-interacting} topological quadrupolar states. We emphasize again here that, though the models are non-interacting, the quantity that we are computing is intrinsically a many-body quantity which requires the full knowledge of the many-body ground state. In the momentum basis, the Hamiltonian for the topological quadrupole insulator is 
\begin{align}
h(\bm{k})& =  \big(\gamma_x + \lambda_x \cos(k_x) \big) \Gamma_4 + \lambda_x \sin(k_x) \Gamma_3 \nonumber \\
&+ \big( \gamma_y + \lambda_y \cos(k_y) \big) \Gamma_2 + \lambda_y \sin(k_y) \Gamma_1 + \delta~\Gamma_0. 
\label{TQI_k_space}
\end{align}
Here, $\Gamma_{a=0,\cdots,4}$ is a proper set of gamma matrices~\cite{suppl_ref}. When $\delta = 0$, there are two anti-commuting  mirror symmetries $\hat{M}_x$ and $\hat{M}_y$ which quantize the quadrupole moment $q_{xy}$ to either 0 or 1/2 mod 1. If $\lambda_x = \lambda_y$ and $\gamma_x = \gamma_y$ are imposed (while keeping $\delta = 0$), there is a $C_4$ symmetry which also quantizes the quadrupole moment $q_{xy}$ to either 0 or 1/2 mod 1. $\delta \ne 0$ breaks both the mirror symmetries and $C_4$ symmetry, which quantize the quadrupoles, but $C_2$ symmetry remains intact which enforce the total polarization to vanish. 

\begin{figure}[t!]
\includegraphics[width=1.0\columnwidth]{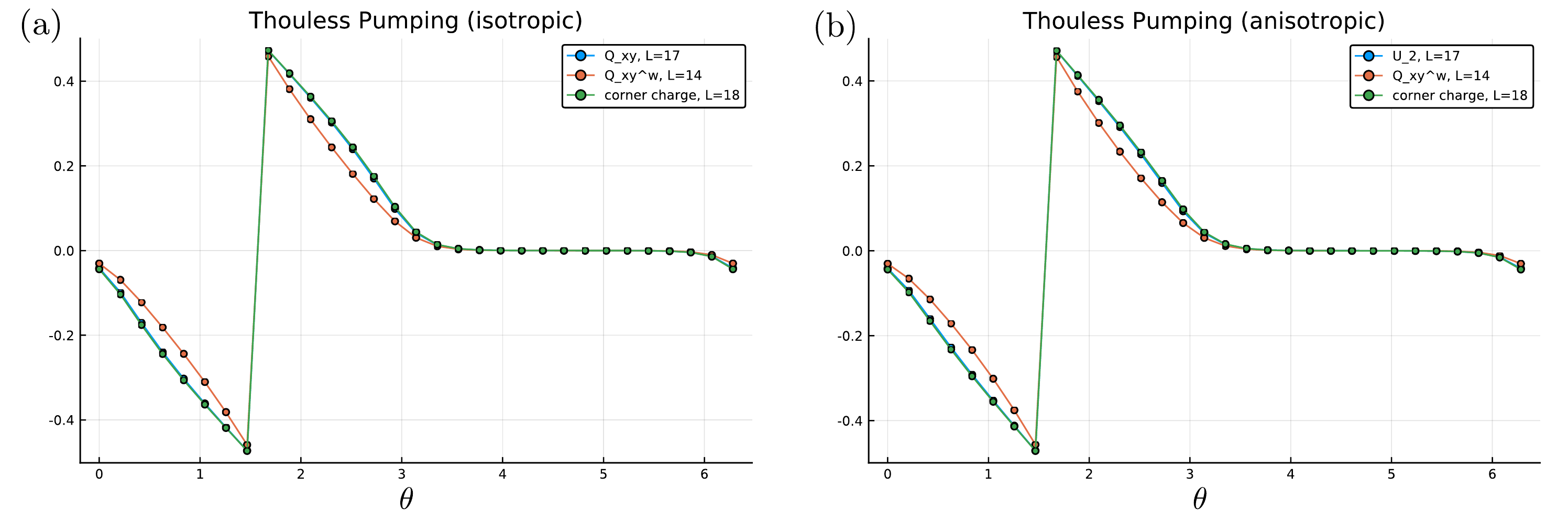}
\caption{Comparison between three different physical quantities, complex phase of $\langle \hat{U}_2 \rangle$, the quadrupole index $Q_{xy}^\omega$, and corner charge $q_\textrm{c}$ for (a) isotropic Thouless pumping Eq.~\eqref{Isotropic} and (b) anisotropic Thouless pumping Eq.~\eqref{Anisotropic}. While the complex phase of $\langle \hat{U}_2 \rangle$ and the corner charge $q_\textrm{c}$ agree with each other with almost no discernible differences, the quadrupole index $Q_{xy}^\omega$ given in terms of the Wannier-sector polarizations agrees with the two only at $\theta = \pi/2, 3\pi/2$, i.e., $\delta = 0$ and the quadrupole moment is quantized.}
\label{Thouless_pumping}
\end{figure}

In FIG.~\ref{U2_TQI_fig}, we present how the complex phase of $\langle \hat{U}_2 \rangle $ changes along different cuts in the parameter space of Eq.~\eqref{TQI_k_space}. The many-body order parameter Eq.~\eqref{Analytic1} reproduces the quantized quadrupole moments as well as the sharp phase transitions between topological and trivial quadrupole insulator, even when the bulk energy gap does not close at the transition point. 

Given the success of our many-body formula in reproducing the phase diagram of Eq.~\eqref{TQI_k_space}, we applied it to the anomalous topological quadrupole insulator\cite{PhysRevB.98.201114}, where the naive application of the nested Wilson loop approach fails to predict the non-zero quantized quadrupole moments and the related phase transition. Here we find remarkably [see FIG.~\ref{U2_TQI_fig} (d)] that the expected quadrupole moment and phase transition are successfully reproduced with our many-body order parameter $\hat{U}_2$.  

To define the quadrupole moment using the nested Wilson loop, the existence of the Wannier gap is crucial as we can then separate a Wannier band to evaluate the Wanner-sector polarization. In the case of Eq.~(\ref{TQI_k_space}) with $\delta=0$, the phase transition between topological and trivial quadrupole insulator happens when the Wannier gap (instead of physical energy gap) closes and re-opens. On the other hand, $|\langle \hat{U}_2 \rangle|$ always vanishes at phase boundary, where the Wannier gap closes i.e., $|\lambda_x/\gamma_x| = 1$ \textit{or} $|\lambda_y/\gamma_y| = 1$ (Note that the physical gap closes only when $|\lambda_x/\gamma_x| = 1$ \textit{and} $|\lambda_y/\gamma_y| = 1$). This highlights the relation between the Wannier gap and $|\langle \hat{U}_2 \rangle|$. 


Having seen the success of $\hat{U}_2 $ operator in detecting quantized quadrupole moment, we next ask what would happen if we break symmetries that protect the quantization of the quadrupole moment. In this case, the quantization of the quadrupole moment is lost and the quadrupole moment can take any value. However, due to the bulk-boundary correspondence~\cite{PhysRevB.48.4442, PhysRevB.95.035421, PhysRevB.96.245115}, we expect that the corner charge is determined by the bulk quadrupole moment and so should be identical to the phase of $\langle \hat{U}_2 \rangle$. For numerical confirmations, we employ two Thouless pumping processes,  one we call isotropic Thouless process and the other we call anisotropic Thouless process, where we relate three different quantities; a) the complex phase factor of $\langle \hat{U}_2 \rangle$ evaluated under full periodic boundary condition, b) the quadrupole index~\cite{PhysRevB.96.245115} $Q_{xy}^\omega = 2 p_x^\omega p_y^\omega$ obtained from the nested Wilson loop, and c) the corner charge $q_\textrm{c}$ localized at one corner under full open boundary condition. We note that the previous study~\cite{PhysRevB.96.245115} found numerically that b) and c) are not equivalent in the absence of the quantizing symmetries.

For the isotropic Thouless pumping case, we change the parameters in Eq.~(\ref{TQI_k_space}) according to
\begin{align}\label{Isotropic}
\begin{cases}
\gamma_x = \gamma_y = 1-0.6 \sin(\theta) \\
\lambda_x = \lambda_y = 1+0.6 \sin(\theta) \\
\delta = 0.6 \cos(\theta)
\end{cases} ,
\end{align}
and for the anisotropic Thouless pumping process, we change the parameters according to
\begin{align}\label{Anisotropic}
\begin{cases}
\gamma_x = 1-0.6 \sin(\theta), \quad \gamma_y = 1-0.5 \sin(\theta) \\
\lambda_x = 1+0.6 \sin(\theta), \quad \lambda_y = 1+0.5 \sin(\theta) \\
\delta = 0.6 \cos(\theta)
\end{cases} ,
\end{align}
where $\theta \in [0, 2\pi]$ is the pumping parameter. In FIG.~\ref{Thouless_pumping}, we have plotted the three fundamentally different physical quantities for (a) isotropic and (b) anisotropic Thouless pumping processes. Apparent from the plots, we see almost no discernible difference between the complex phase factor of $\langle \hat{U}_2 \rangle$ and the corner charge $q_\textrm{c}$, indicating that the many-body order parameter faithfully represents the physical quadrupole moment. On the other hand, the Wilson loop index $Q_{xy}^\omega$ differs from the two in general and agrees with the two only at quantized values. This is the numerical proof that $\hat{U}_2$ can measure the physical quadrupole moment density in a crystal.  

\begin{figure}[t!]
\includegraphics[width=1.0\columnwidth]{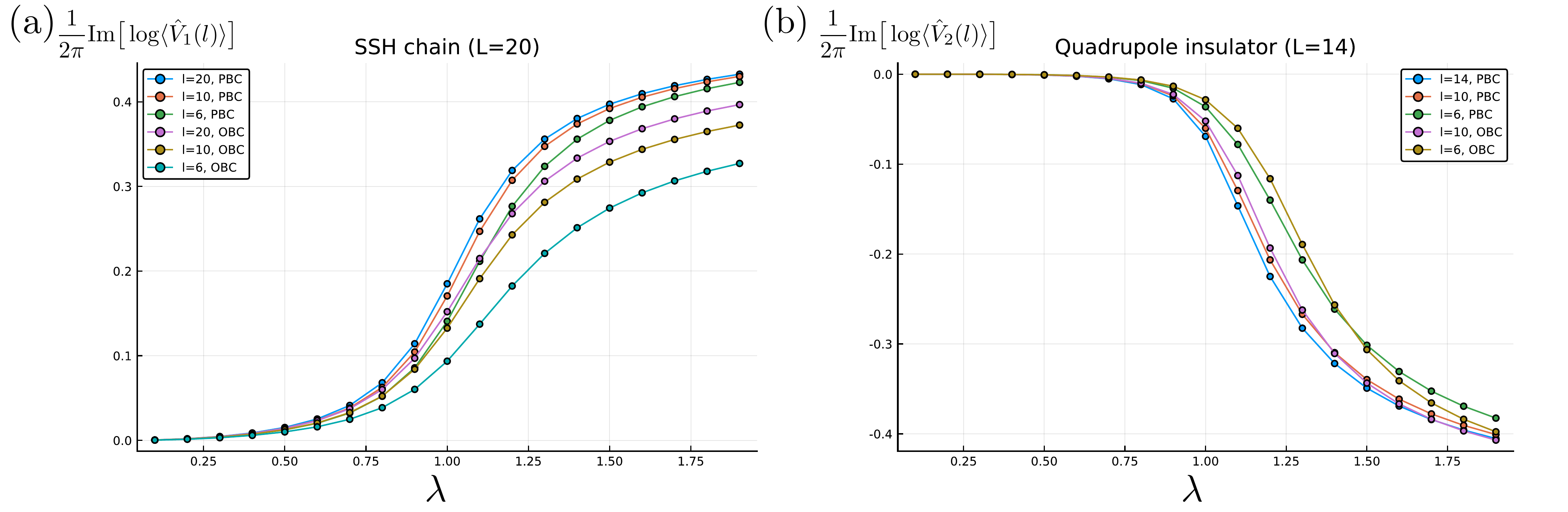}
\caption{Evaluation of the phase of $\langle \hat{V}_{a=1,2} (l) \rangle $ of Eq.~\eqref{Analytic_G1} and Eq.~\eqref{Analytic_G2} for various values of $l$ in the case of (a) Su-Schrieffer-Heeger chain and (b) Quadrupole insulator Eq.~(\ref{TQI_k_space}). We set $\gamma=1.0$ ($\gamma_x = \gamma_y = 1.0$) and $\delta = 0.2$ and change $\lambda$ ($=\lambda_x = \lambda_y$).} 
\label{Partial}
\end{figure}

Next, we perform the numerical check for the generalized many-body order parameters Eq.~\eqref{Analytic_G1} and Eq.~\eqref{Analytic_G2} in FIG.~\ref{Partial}. We compute the phases of these invariants for various $l$'s for the Su-Schrieffer-Heeger chain~\cite{PhysRevLett.42.1698} and topological quadrupolar insulators Eq.~(\ref{TQI_k_space}), which are both subject to the nonzero $\delta$ that breaks the quantization of polarization and quadrupole moments. For the Su-Schrieffer-Heeger chain, we have used 
\begin{equation}
h_\textrm{SSH} (k) = \left(
\begin{array}{cc}
\delta & \gamma+\lambda e^{-ik} \\
\gamma + \lambda e^{ik} & -\delta
\end{array} \right) \nonumber .
\end{equation}
Surprisingly, the overall trend is insensitive to which boundary condition is used and insensitive even when $l$ is less than a half of the system size. Generically, we find that as $l$ approaches the full system size and as the system size becomes large enough, these invariants collapse to the corner charge as we predicted. 

To benchmark our formula for the interacting models, we also confirmed that our many-body order parameter can diagnose the non-trivial topology of the exactly-solvable bosonic higher-order topological insulator\cite{Hughes} with the corner spin.\cite{suppl_ref}   

Finally, we also have applied our invariant Eq.~\eqref{Analytic1} to another model, the ``boundary polarization model"~\cite{PhysRevB.96.245115} supporting quantized corner charges $q_c = \pm 1/2$ yet having vanishing quadrupole moment (with suitable perturbations). Our order parameter correctly captures the vanishing quadrupole moment despite of the nonzero corner charges coming from boundary-alone polarizations.\cite{suppl_ref} We also find several interesting numerical observations on $\hat{U}_a$ including the dependence on coordinate parametrizations, which we discuss in detail in supplemental materials.\cite{suppl_ref}

\textbf{3. Conclusions}: We have proposed many-body order parameters which work for both the quantized and non-quantized multipoles in crystalline many-body quantum states. Using analytical arguments, we have demonstrated that they can be identified with the corner charges originating from the multipole moments. Based on the reasonings of the effective field theories, we have provided an extension of our formulas, which work even for the case of open boundary condition. Finally, we provided numerical proof of our order parameters by presenting explicit computations in toy models and confirmed that they can measure the multipole moments. Compared to the nested Wilson loop indices~\cite{PhysRevB.96.245115}, our order parameters can measure the multipole moments even in the absence of quantizing symmetries and can be applied to interacting systems as well. On the other hand, in the presence of the quantizing symmetries, we have consistent results with the nested Wilson loop indices and provide complementary real-space formulation.

\acknowledgements
We thank Akira Furusaki, Jung Hoon Han, Hyun Woong Kwon, Joel E. Moore, Masaki Oshikawa, Kwon Park, Shinsei Ryu, and Mike Zaletel for helpful discussions and comments. KS is supported by PRESTO, JST (JPMJPR18L4) and GYC is supported by BK 21 plus program at POSTECH in Korea. GYC thanks KIAS in Korea and RIKEN in Japan for their hospitality where some part of this work has been done. We thank Taylor Hughes for informing us about their parallel independent work\cite{WWH} and sharing it with us before uploading to arXiv. When their work and our work overlap, they agree with each other. We also thank Andrei Bernevig and Ben Wieder for their comments and explanation on nested Wilson loop indices. 

\bibliography{Multipole-Resta-Formula}

\begin{thebibliography}{35}
\bibitem [{com()}]{comment1}%
  \BibitemOpen
  \href@noop {} {}\bibinfo {note} {For the systems with even $L_x \cdot L_y$,
  there is an additional minus sign (which can also be noticed in Resta's
  original paper~\cite{PhysRevLett.80.1800}) but the similar discussion can be
  straightforwardly made by noting that the topological part of the quadrupole
  moments are defined as the relative deviation from that of the trivial
  insulators. In the remaining text, we always take into account this
  (possible) additional $(-1)$ sign factor when evaluating $\hat{U}_a$. Note
  that the sign factor can also be computed by evaluating $\hat{U}_a$ for
  background positive charge distributiuon.}\BibitemShut {Stop}%
\end{thebibliography}

\end{document}


\title{Supplemental Materials for ``many-body order parameters for Multipoles in Higher-Order Topological Insulators"}

\author{Byungmin Kang}         
\affiliation{School of Physics, Korea Institute for Advanced Study, Seoul 02455, Korea}

\author{Ken Shiozaki}         
\affiliation{Yukawa Institute for Theoretical Physics, Kyoto University, Kyoto 606-8502, Japan}

\author{Gil Young Cho}         
\affiliation{Department of Physics, Pohang University of Science and Technology (POSTECH), Pohang 37673, Republic of Korea}

\date{\today}
\maketitle
\tableofcontents
\appendix
\section{Details on Symmetry Properties of $\hat{U}_2$}
In this section, we present the details of the symmetry actions on the many-body order parameters in $\hat{U}_2$. This highlights under what conditions $\hat{U_2}$ is well-defined and yields a quantized value when evaluated with respect to a many-body state. In the following, we discuss the action of the crystalline $C_4$ symmetry on the many-body order parameter and the large translation symmetry.  

\subsection{Large Translation Symmetry}
Here we provide the details of the large coordinate translation operation $x \to x + L_x$ (with $L_x$ being the system size along $x$-direction) on the many-body quadrupolar invariant Eq. (2) in main text. As noted in the main text, we find 
\begin{align}
x \to x+L_x:\quad \hat{U}_2 \to \hat{U}_2 \hat{U}_{1;y}
\end{align}
in which $\hat{U}_{1;y}$ is the many-body measure for the polarization along $y$ direction. Now the key step that we used in the main text is  
\begin{align}
\langle \hat{U}_2 \rangle &= \langle  \hat{U}_2 \rangle \langle \hat{U}_{1;y} \rangle + \mathcal{O}\Big(\frac{\epsilon_0}{E_{\text{gap}}}\Big), \nonumber\\ 
&=\langle  \hat{U}_2 \rangle \langle \hat{U}_{1;y} \rangle + \mathcal{O}\Big(\frac{\xi}{a_0}\Big), 
\end{align}
where $\epsilon_0$ and $a_0$ are the unit energy and unit scale which make the expressions dimensionless. From the first line to the second line, we used the fact that $\frac{1}{E_{\text{gap}}} \propto \xi $ with $\xi$ being the correlation length. When the correlation length becomes an atomic scale $\xi \to 0^+$ (ultra-short correlated), i.e., the ideal atomic insulator limit, we see that $\langle \hat{U}_2 \rangle = \langle \hat{U}_2 \rangle \langle \hat{U}_1 \rangle $ which enforces $\langle \hat{U}_1 \rangle = 1$. 

The above equations can be derived from the following observation: 
\begin{align}
\hat{U}_1 |GS\rangle = \langle \hat{U}_1 \rangle |GS \rangle +  \mathcal{O}\Big(\frac{\epsilon_0}{E_{\text{gap}}}\Big),
\end{align}
which is a slight rewriting of Eq. (37) in Ref.~[\onlinecite{Watanabe-Oshikawa}]. 

\subsection{$C_4$ Symmetry and Mirrors}
With the discussion above, we now discuss the action of $C_4$ on $\hat{U}_2$ for the domain $\mathcal{M} = [1, L_x ] \times [1, L_y]$. Obviously the system respects the $C_4$ symmetry only when $L_x = L_y =L$. Furthermore, it is not difficult to see that the domain $\mathcal{M} = [1, L] \times [1, L]$ is transformed into $\mathcal{M}' = [-L, -1] \times [1, L]$. Hence, we find that 
\begin{align}
C_4: \hat{U}_2 \to \hat{U}_2^* \hat{U}_{1;x}. 
\end{align}
Since, in the ideal limit $\xi \to 0^+$, $\langle \hat{U}_2^* \hat{U}_{1;x} \rangle = \langle \hat{U}_2^* \rangle \langle \hat{U}_{1;x} \rangle = \langle \hat{U}_2^* \rangle $ (with vanishing polarization) and thus $\langle \hat{U}_2 \rangle  = \langle \hat{U}_2^* \rangle$. This enforces $Q_{xy} = -Q_{xy}$ which is consistent with the previous work\cite{Benalcazar61}. In fact, even for the mirror symmetries $M_x \times M_y$ discussed in the main text, extra factors of $\hat{U}_{1;\mu = x,y}$'s do show up because of the transformations on the domain $\mathcal{M}$, which need to be included. Nevertheless, the conclusion made in the main text remains the same, i.e., $C_4$ or mirrors flip the sign of $Q_{xy}$, which can be explicitly seen in our many-body order parameters in an appropriate limit.

\section{Details on Effective Field Theory of Quadrupole Moments}
In this section, we provide the details of the effective field theory for the quadrupolar insulators. We first derive the effective response from the semi-classical picture and then use it to reproduce the quitessential features\cite{PhysRevB.96.245115} of the topological quadrupolar insulators. Furthermore, we also show the link between the ground state expectation value $\langle \hat{U}_a \rangle$ and the partition function. Although we mainly focus on the quadrupolar case, let us remark that all the discussions here can be straightforwardly generalized to  the octupolar and higher-polar cases.  

\begin{figure}[!ht]
\centering\includegraphics[width=0.7\textwidth]{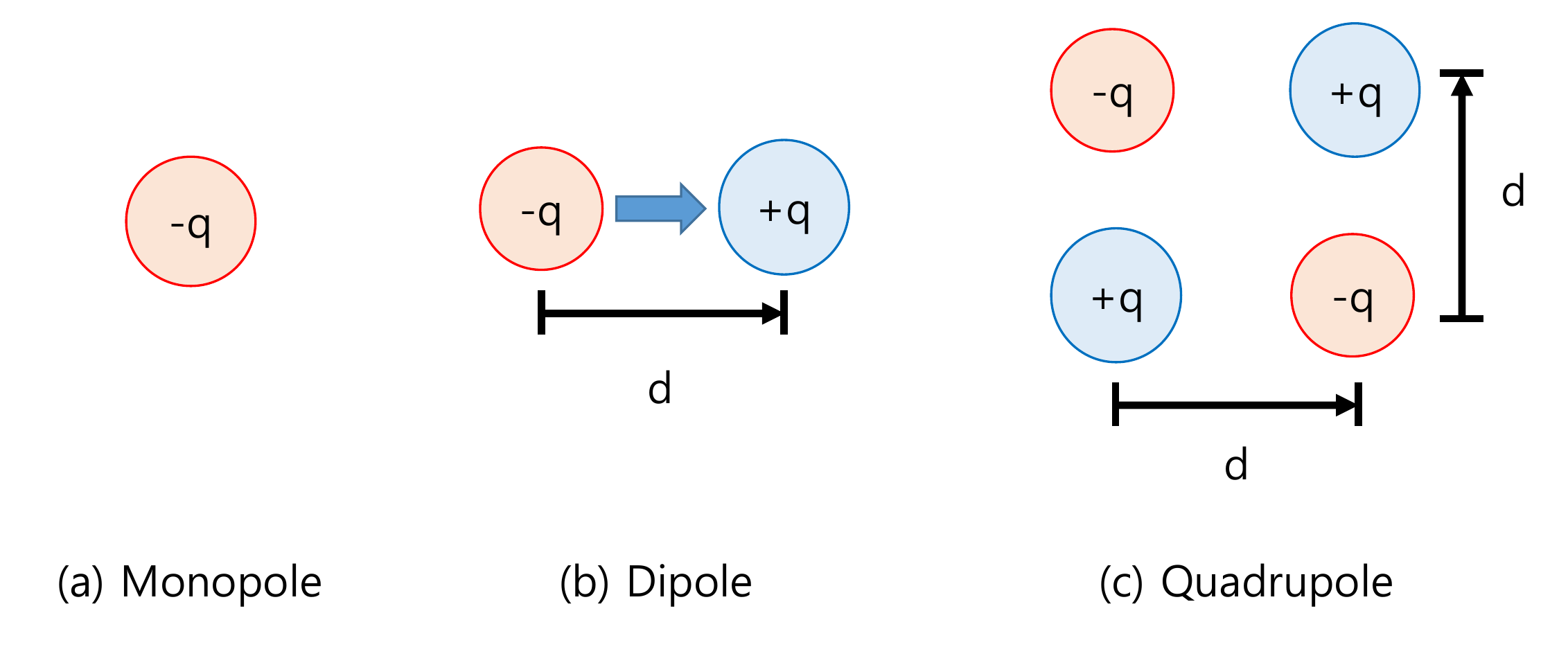}
\caption{Charge configuration in real space with multipole moments}
\label{Multipole}
\end{figure}

\subsection{Derivation of Effective Action: Multipolar Electromagnetism}
We start from the effective responses for electric multipoles. First, the uniform monopole or charge density $\rho$ in a spatial region $\mathcal{M}$ has the following response: 
\begin{align}
S_{eff} = \int_{\mathcal{M}} d\tau d^{d}r ~ \rho V(\bm{r})
\end{align} 

Second, a single dipole at the site $\bm{r}$ with the magnitude $\vec{P} = q\bm{d}$ has the following response: 
\begin{align} 
S_{\text{dipole}} = \int d\tau ~q \Big[V(\bm{r}+\bm{d}) - V(\bm{r})\Big] \approx \int d\tau ~ q \bm{d}\cdot \bm{\partial} V(\bm{r})  =\int d\tau ~ \vec{P} \cdot \vec{E}.
\end{align} 
Hence, when there are uniform polarization density $\vec{P}$ over the area $\mathcal{M}$, we find 
\begin{align}
S_{eff} = \int_{\mathcal{M}} d\tau d^{d}r ~ \vec{P}\cdot \vec{E}. 
\end{align} 

Third, a single quadrupole $Q_{xy} = q d^2$ at the site $\bm{r} = (x,y) $ (see Fig.~\ref{Multipole}) has the following response: 
\begin{align} 
S_{\text{quadrupole}} &= \int d\tau ~q \Big[V(x,y) - V(x+d, y) - V(x, y+d) +V(x+d, y+d) \Big] \approx \int d\tau ~ q d^2 \partial_x \partial_y V(x,y)  \nonumber\\ 
&=\int d\tau ~ \frac{Q_{xy}}{2} \Big[\partial_x E_y + \partial_y E_x \Big] .
\end{align} 
Thus, when there are uniform quadrupole density $Q_{xy}$ over the area $\mathcal{M}$, we find 
\begin{align}
S_{eff} = \int_{\mathcal{M}} d\tau d^{d}r ~ \frac{Q_{xy}}{2} \Big[\partial_x E_y + \partial_y E_x \Big] , 
\label{S_eff_quad} 
\end{align} 
which is the effective action used in the main text. 

It is not difficult to show that the uniform octupole density $O_{xyz}$ over the region $\mathcal{M}$ has the following response:  
 \begin{align}
S_{eff} = \int_{\mathcal{M}} d\tau d^{d}r ~ \frac{O_{xyz}}{3} \Big[\partial_x \partial_y E_z + \partial_z \partial_y E_x  +\partial_z \partial_x E_y  \Big].  
\end{align} 

\subsection{Corner Charge and Thouless Pumping}
From the effective action for the quadrupole Eq.~\eqref{S_eff_quad}, we can reproduce the essential features\cite{PhysRevB.96.245115} of the topological quadrupolar insulators. For this, let us consider a rectangular region $\mathcal{M} = (-L_x , L_x) \times (-L_y, L_y) $ with uniform  $Q_{xy} \neq 0$, which has an open boundary with the vacuum $Q_{xy} = 0$ as described in Fig.~\ref{Quadrupole1} (a). 

\subsubsection{Corner charge}
We first compute the corner charge. It can be done by computing the equation of motion for $A_0$ in the effective response. 
\begin{align}
\rho (\bm{r})= \frac{\delta \mathcal{L}_{eff}}{\delta A_0 (\bm{r})} = \partial_x \partial_y Q_{xy}
\end{align}
Obviously, the RHS of the above equation, $\partial_x \partial_y Q_{xy}$, is non-zero only at the corners of $\mathcal{M}$. Hence, we find  
\begin{align}
\rho(\bm{r}) = Q_{xy} \Big[\delta(x-L_x)\delta(y-L_y)  - \delta(x + L_x)\delta(y-L_y) - \delta(x - L_x)\delta(y + L_y) +\delta(x + L_x)\delta(y + L_y) \Big]. 
\end{align}
This is consistent with the intuitive picture of the corner charge generated from the quadrupolar moment density. Integrating over the quadrant of the $\mathcal{M}$, we finally find that 
\begin{align}
q_c = \pm Q_{xy},  
\end{align}
whose sign depends on which quadrant of the space that we integrate over. 
  
\subsubsection{Thouless pumping}
\begin{figure}[t!]
\centering\includegraphics[width=0.7\textwidth]{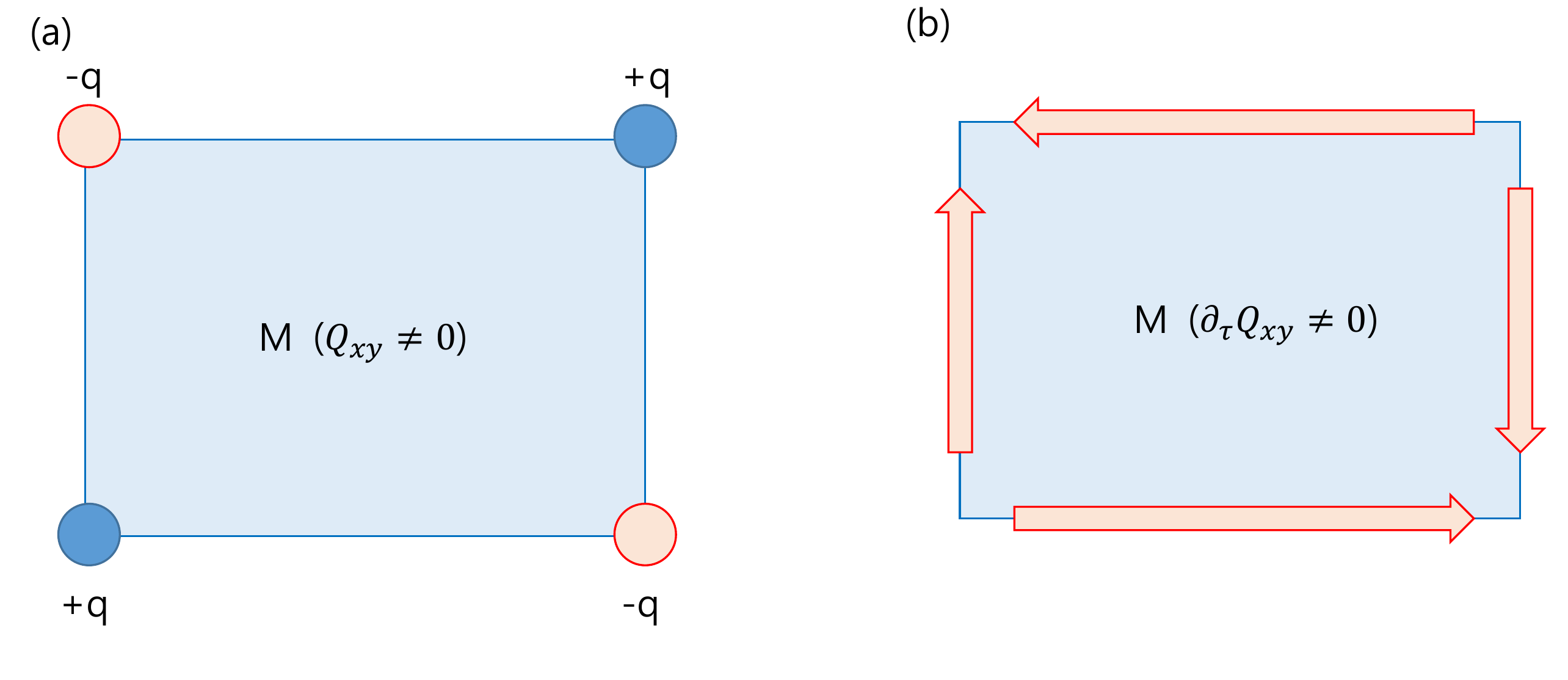}
\caption{(a) Corner charges of quadrupolar insulator derived from the effective action. Charges are only localized at four corners of the rectangular region $\mathcal{M}$, which has nonzero quadrupole moment $Q_{xy}=0$. Note that the vacuum has trivial quadrupole moment $Q_{xy}=0$. (b) Charge current configuration derived from the effective action under the Thouless pumping. In this case, charges flow only along the boundary of $\mathcal{M}$ and their directions are denoted as arrows.}
\label{Quadrupole1}
\end{figure}

Next we imagine that the uniform quadrupolar moment density $Q_{xy}$ depends on time, i.e., we have $Q_{xy} = Q_{xy}(\tau)$. This allows the charge currents to flow along the boundary of the topological quadrupolar insulators. To see this, we calculate 
\begin{align}
J_x = \frac{\delta \mathcal{L}_{eff}}{\delta A_x} =- \partial_y \partial_\tau Q_{xy}, ~~  J_y = \frac{\delta \mathcal{L}_{eff}}{\delta A_y} = -\partial_x \partial_\tau Q_{xy},
\end{align}
which is essentially equivalent to the charge flow in the process of Thouless pumping in the topological quadrupolar insulator, obtained in Ref.~[\onlinecite{PhysRevB.96.245115}]. See FIG.~\ref{Quadrupole1} (b) for the direction of current along the boundary of $\mathcal{M}$ during the Thouless pumping.

\subsection{From $\langle GS| \hat{U}_a |GS \rangle$ to Partition Functions}
Here we relate the ground state overlaps and the partition functions, which has been extensively used in many-body order parameters for symmetry-protected topological states \cite{PhysRevB.91.195142, PhysRevLett.118.216402}  on the relation of the ground state expectation values of the symmetry operators and the partition functions. This is in fact a small variation of the standard Dyson formula, which can be found in any quantum field theory text books. 

\begin{figure}[h!]
\centering\includegraphics[width=0.7\textwidth]{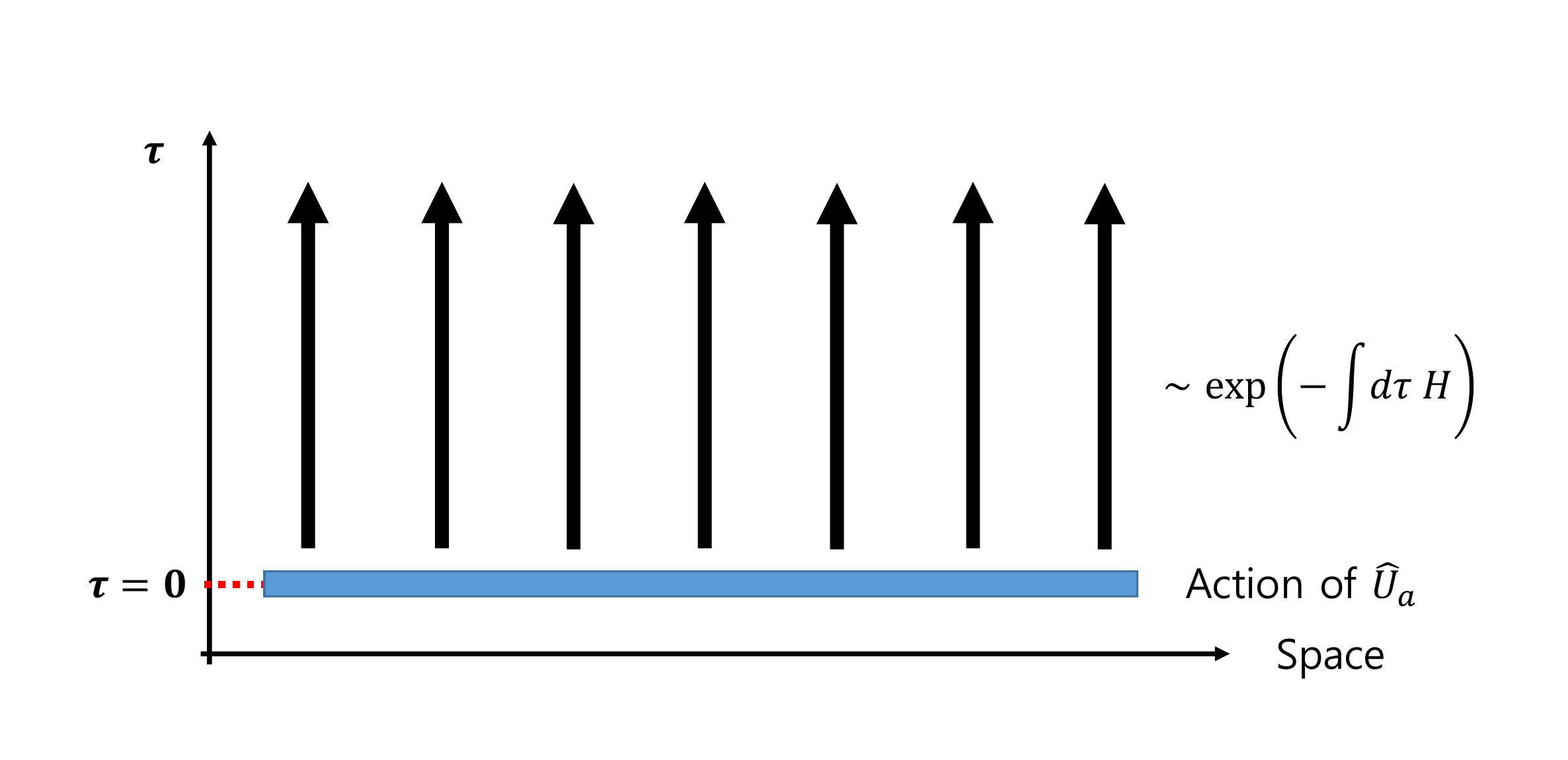}
\caption{Interpretation of partition functions with the insertion of $\hat{U}_a$. $\hat{U}_a$ acts at $\tau=0$ followed by the imaginary time evolution along $\tau$. Periodic boundary condition is used along $\tau$ direction.}
\label{PartitionFunction}
\end{figure}

We first start with noting that  
\begin{align}
|GS \rangle = \frac{1}{\sqrt{Z}} \sum_n e^{-\frac{\beta}{2} H} |n\rangle, ~~ Z = \text{Tr} [ e^{-\beta H} ], 
\end{align}
where the true ground state can be obtained by letting the ``temperature" $1/\beta \to 0$. This implies that 
\begin{align}
\langle GS| \hat{U}_a |GS \rangle  \propto \frac{1}{Z} \sum_n \langle n | e^{-\frac{\beta}{2} H} \hat{U}_a e^{-\frac{\beta}{2} H}  |n \rangle  = \frac{1}{Z} \text{Tr} \Big[e^{-\beta H} \hat{U}_a\Big].  
\label{U_2_to_part}
\end{align}
Now, in order to relate Eq.~\eqref{U_2_to_part} to the effective functional against the gauge potential configuration set by $\hat{U}_a$, we need to properly interpret the partition function in the RHS of Eq.~\eqref{U_2_to_part}. Pictorially, this can be represented as the imaginary time evolution (See Fig.~\ref{PartitionFunction}). For a proper interpretation, let us break the partition function into the two parts. The first part is the action of $\hat{U}_a$. Note that these $\hat{U}_a$'s involve complex phase factors proportional to the particle number, i.e., $\hat{U}_a \sim \exp\Big[i \phi (\bm{r}) \hat{n}(\bm{r}) \Big]$ (with $\phi(\bm{r})$ depending on $a$). This means that $\hat{U}_a$ appearing on the partition function Eq.~\eqref{U_2_to_part} is the same as the action of the gauge potential $A_0 (\bm{r}, \tau)$ on the states at $\tau =0$, i.e., $A_0(\bm{r}, \tau)\propto \delta(\tau)$. In the second part, we act $\exp(-\beta H)$, which is the imaginary time-evolution on the states after the action of $\hat{U}_a$.  

Because the RHS of Eq.~\eqref{U_2_to_part} can be interpreted as the evolution of the states under the  gauge potentials, the RHS should be equivalent to the following: 
\begin{align}
\langle GS| \hat{U}_a |GS \rangle  \sim |Z| \exp\Big(iS_{eff}[A_0 (\bm{r}, \tau)]\Big), 
\end{align} 
in which $S_{eff}[A_0 (\bm{r}, \tau)]$ is the effective response against $A_0(\bm{r}, \tau)$, which is set by $\hat{U}_a$. Here, we also used the fact that the multipolar electromagnetic responses are \textit{topological} (or Berry phase) because they involve a single time-derivative $\sim \partial_\tau$. Thus, this multipolar response against $A_0 (\bm{r}, \tau)$ contributes to the imaginary part of the $\langle \hat{U}_a \rangle$. With these, it is straightforward to confirm that our many-body order parameters are designed to saturate $S_{eff} = 2\pi Q_{xy}$ for the quadrupole case and $2\pi P_x$ for the dipole case. 

\section{Exactly-Solvable Bosonic Higher-Order Topological Insulator}
Here we prove that our many-body order parameter can diagnose the bosonic higher-order topological insulator\cite{Hughes} at the exactly-solvable points, i.e., where the correlation length is the size of the unit cell. The model that we study here has four spin-$\frac{1}{2}$'s per unit cell and it consists of the two-spin interactions respecting $SU(2)$.  
\begin{align}
H = J \sum_{\text{P}} \sum_{\langle i, j \rangle \in \text{P}} \hat{S}_i \cdot \hat{S}_j  + \lambda \sum_{\bm{R}} \sum_{\langle i, j \rangle \in \bm{R}} \hat{S}_i \cdot \hat{S}_j 
\end{align}
Here $J$ parametrizes the strength of the spin-$\frac{1}{2}$'s if the two spins $i, j$ belong to the \textit{intersite} plaquette and $\lambda$ parametrizes the strength of the spin-$\frac{1}{2}$'s if the two spins $i,j$ belong to the \textit{intrasite} plaquette. Obviously $S_z$ is conserved and hence we have $U(1)$ symmetry. 

If $\lambda =0$ or $J=0$, the model is exactly solvable and the ground state can be explicitly written out\cite{Hughes}. At these limits, the ground state is the simple product of the wavefunctions per each four spin-$\frac{1}{2}$'s connected by either by $J$ or $\lambda$: 
\begin{align}
|P\rangle = \frac{1}{2\sqrt{2}}\Big[ |\uparrow \uparrow \downarrow \downarrow \rangle +|\downarrow \uparrow \uparrow  \downarrow \rangle + |\uparrow \downarrow \downarrow \uparrow \rangle \Big] - \frac{1}{2} \Big[|\uparrow \downarrow \uparrow  \downarrow \rangle +|\downarrow \uparrow  \downarrow \uparrow  \rangle \Big]. 
\end{align}
The many-body order parameter can be constructed as following: 
\begin{align}
\hat{U}_2 =  \exp \Big[ \frac{2\pi i}{L_x L_y} \sum_{\bm{r}}xy (\hat{S}_{\bm{r}}^z - \bar{S})\Big],
\end{align}
where $\bar{S}=2$ is introduced. This $\bar{S}$ is the average background spin density, which we subtract out. This can be explicitly computed for the exact ground states, and we obtain that the topological state has $(-1) = \langle \hat{U}_2 \rangle $ (for $J\neq 0$ and $\lambda =0$) and $(+1) = \langle \hat{U}_2 \rangle $ (for $J =0$ and $\lambda \neq 0$).

\section{Details on Tight-Binding Models}
In this section, we summarize the physical properties and the symmetries of the tight-binding models for the topological insulator appearing in the main text.

\subsection{Su-Schrieffer-Hegger chain}
A minimal model for one-dimensional crystalline topological insulator is the Su-Schrieffer-Hegger (SSH) chain~\cite{PhysRevLett.42.1698} which is a tight-binding model having alternating ``weak'' and ``strong'' hopping terms:
\begin{equation}
H_\textrm{SSH} = \sum_{r} \Big[ \big(\gamma c_{r,1}^\dagger c_{r,2} + \lambda c_{r,2}^\dagger c_{r+1,1} + \textrm{h.c.} \big) + \delta \big(c_{r,1}^\dagger c_{r,1} - c_{r,2}^\dagger c_{r,2} \big) \Big] ,
\end{equation}
where $c_{r,\alpha=1,2}^\dagger$ is the electron creation operator at site $r$ orbital $\alpha$, $\gamma$ and $\lambda$ are inter- and intra-site hopping strengths, and $\delta$ is the onsite potential strength. Under the periodic boundary condition, it is convenient to work with the Bloch basis:
\begin{equation}
c_{k, \alpha}^\dagger = \frac{1}{\sqrt{L}} \sum_{r=1}^L e^{-ikr} c_{r,\alpha}^\dagger,
\end{equation}
where $k = \frac{2\pi n}{L}$ for some $n \in \mathbb{Z}$, $\alpha = 1,2$ is the orbital index, and $L$ is the system size. $c_{k, \alpha}$ can be defined analogously. The block Hamiltonian in terms of the Bloch state is given by
\begin{equation}
h_\textrm{SSH} (k) = \left(
\begin{array}{cc}
\delta & \gamma + \lambda e^{ik} \\
\gamma + \lambda e^{-ik} & -\delta
\end{array}
\right) .
\label{suppl_Ham_SSH_k}
\end{equation}

\subsection{Topological Quadrupole insulator}
A minimal model for topological quadrupole insulator~\cite{PhysRevB.96.245115}, is given by
\begin{align}
H = \sum_{\r} \Big[ &\gamma_x \big(c_{\r, 1}^\dagger c_{\r, 3} + c_{\r, 2}^\dagger c_{\r, 4} + \textrm{h.c.} \big) + \gamma_y \big( c_{\r, 1}^\dagger c_{\r, 4} - c_{\r, 2}^\dagger c_{\r, 3} + \textrm{h.c.} \big) + \lambda_x \big( c_{\r, 1}^\dagger c_{\r+\hat{x}, 3} + c_{\r, 4}^\dagger c_{\r+\hat{x}, 2} + \textrm{h.c.} \big) \nonumber \\
& + \lambda_y \big( c_{\r, 1}^\dagger c_{\r + \hat{y}, 4} - c_{\r, 3}^\dagger c_{\r + \hat{y}, 2} + \textrm{h.c.} \big) + \delta \big(c_{\r, 1}^\dagger c_{\r, 1} + c_{\r, 2}^\dagger c_{\r, 2} - c_{\r, 3}^\dagger c_{\r, 3} - c_{\r, 4}^\dagger c_{\r, 4} \big) \Big] ,
\label{suppl_Ham_TQI}
\end{align}
where $c_{\r, \alpha}^\dagger$ and $c_{\r, \alpha}$ are the fermion creation and annihilation operator at site $\r = (x,y)$ and $\alpha =1,2,3,4$ is the orbital index. If we impose the periodic boundary condition along $x$- and $y$-direction then it is convenient to work with the Bloch (momentum) basis
\begin{equation}
c_{\k, \alpha}^\dagger = \frac{1}{\sqrt{L_x L_y}} \sum_{x=1}^{L_x} \sum_{y=1}^{L_y} e^{-i \k \cdot \r} c_{\r, \alpha}^\dagger , \nonumber
\end{equation}
where $\k = (k_x, k_y) = (\frac{2\pi n_x}{L_x}, \frac{2\pi n_y}{L_y})$ for some $n_x, n_y \in \mathbb{Z}$ is the Bloch momentum, $\alpha =1,2,3,4$ is the orbital index, and $L_x$ and $L_y$ is the system size along $x$- and $y$-direction. $c_{\k, \alpha}$ is defined analogously. Using the Bloch basis, the Hamiltonian Eq.~\eqref{suppl_Ham_TQI} takes the block diagonal form
\begin{equation}
h(\k) = \big( \gamma_x + \lambda_x \cos (k_x) \big) \Gamma_4 + \lambda_x \sin (k_x) \Gamma_3 + \big( \gamma_y + \lambda_y \cos(k_y) \big) \Gamma_2 + \lambda_y \sin(k_y) \Gamma_1 + \delta \Gamma_0,
\label{suppl_Ham_TQI_k}
\end{equation}
where $\Gamma_0 = \tau_3 \otimes \tau_0$, $\Gamma_i = -\tau_2 \otimes \tau_i$ for $i=1,2,3$, $\Gamma_4 = \tau_1 \otimes \tau_0$, where $\tau_0$ is the $2 \times 2$ identity matrix and $\tau_i$ is the $i$th Pauli matrix. When $\delta = 0$, the physical energy band spectrum is gapless only when $|\gamma_x/\lambda_x| = 1$ and $|\gamma_y/\lambda_y| = 1$ while the Wannier gap closes when $|\gamma_x/\lambda_x| = 1$ or $|\gamma_y/\lambda_y| = 1$. The ground state at half-filling becomes a topological quadrupole insulator when $\delta = 0$, $|\gamma_x/\lambda_x| < 1$, and $|\gamma_y/\lambda_y| < 1$.

\subsubsection{Symmetries of the tight-binding model}
When $\delta = 0$ in Eq.~\eqref{suppl_Ham_TQI}, there exists various symmetries which quantize the quadrupole moment $q_{xy} = 0, 1/2 \mod 1$. In the following, we summarize the symmetries of Eq.~\eqref{suppl_Ham_TQI} in the case of $\delta = 0$. 

\subsubsection{Mirror symmetries}
With respect to the Bloch basis, $x$- and $y$-mirror symmetry can be written as
\begin{equation}
\hat{M}_x = i \tau_1 \otimes \tau_3 \quad \textrm{and} \quad \hat{M}_y = i \tau_1 \otimes \tau_1,
\end{equation}
and the block Hamiltonian Eq.~\eqref{suppl_Ham_TQI_k} transforms as
\begin{equation}
\hat{M}_x h(k_x, k_y) \hat{M}_x^{-1} = h(-k_x, k_y) \quad \textrm{and} \quad \hat{M}_y h(k_x, k_y) \hat{M}_y^{-1} = h(k_x, -k_y) .
\end{equation}
Because of the $\pi$ flux threaded in each plaquette, two mirror symmetries do not commute but anticommute: $\{\hat{M}_x, \hat{M}_y\} = 0$. 

\subsubsection{$C_4$ symmetry}
When $\lambda_x = \lambda_y$ and $\gamma_x = \gamma_y$ in addition to $\delta = 0$, there exists $C_4$ symmetry which also quantizes $q_{xy}$. With respect to the Bloch basis, the $C_4$ symmetry can be represented as
\begin{equation}
\hat{r}_4 = \left(
\begin{array}{cc}
0 & \tau_0 \\
-i \tau_2 & 0
\end{array}
\right)
\end{equation}
and also
\begin{equation}
\hat{r}_4 h(k_x, k_y) \hat{r}_4^{-1} = h(k_y, -k_x)
\end{equation}
holds. 

\subsubsection{$C_2$ symmetry}
Even when $\delta \ne 0$, there exists $C_2$ symmetry which quantizes the polarization in $x$- and $y$-direction. So when $|\delta|$ is small compared to other parameters in Eq.~\eqref{suppl_Ham_TQI_k}, both $x$ and $y$ polarization are 0 hence the quadrupole moment $q_{xy}$ is well-defined although it is no-longer quantized. With respect to the Bloch basis, the $C_2$ operator can be represented as 
\begin{equation}
\hat{r}_2 = - i \tau_0 \otimes \tau_2,
\label{suppl_C2_k}
\end{equation}
and also
\begin{equation}
\hat{r}_2 h(\k) \hat{r}_2^{-1} = h(-\k)
\end{equation}
holds. 

\subsubsection{Symmetry breaking perturbations}
Here, we present how onsite symmetry breaking perturbations affect the bulk quadrupole moment defined as a complex phase of $\langle \hat{U}_2 \rangle$ when perturbing away from a topological quadrupole insulator point. In adding perturbations, we still want to keep $C_2$ symmetry as this would enforce the quantization of polarization in $x$- and $y$-direction, i.e., total polarization remains zero when the perturbation is small.

Starting from the $C_2$ operator Eq.~\eqref{suppl_C2_k}, one can classify $4 \times 4$ matrices that commute with $\hat{r}_2$. The general form of such matrix can be expressed as
\begin{equation}
H_\textrm{pert} = \left(
\begin{array}{cc}
a_r \tau_0 + c_r \tau_2 & e_c \tau_0 + f_c \tau_2  \\
e_c^* \tau_0 + f_c^* \tau_2 & b_r \tau_0 + d_r \tau_2
\end{array} \right) ,
\label{C2_pert}
\end{equation}
where $a_r, b_r, c_r, d_r$ are real parameters, $e_c$ and $f_c$ are complex parameters, and asterisk denotes the complex conjugation. One can immediately see that $\delta$ term in Eq.~(\ref{suppl_Ham_TQI_k}) is reproduced when $a_r = -b_r = \delta$ and set all the other coefficients to zero in Eq.~(\ref{C2_pert}). Let us also note that when ($a_r = b_r = 0$, $c_r = -d_r$, $e_c = -i f_c \in \mathbb{R}$), Eq.~(\ref{C2_pert}) preserves two mirror symmetries and when ($a_r = b_r = 0$, $c_r = d_r$, $e_c = -i f_c \in \mathbb{R}$), Eq.~(\ref{C2_pert}) preserves $C_4$ symmetry. 

One can check numerically that when the perturbation Eq.~(\ref{C2_pert}) is \textit{onsite}, only nonzero $a_r$ and $b_r$ do change the quadrupole moment whereas other terms does not. When $a_r$ and $b_r$ are nonzero, we can always add an identity matrix with some coefficient so that the perturbation reduces to the $\delta$ term in Eq.~(\ref{suppl_Ham_TQI_k}). Hence Eq.~(\ref{suppl_Ham_TQI_k}) is quite generic. 

\subsection{Topological Octupolar Insulator}

A minimal model for an octupole insulator is given by~\cite{PhysRevB.96.245115}
\begin{align}
h_\textrm{octupole} (\k) = &\lambda_y \sin(k_y) \Gamma_1' + \big[ \gamma_y + \lambda_y \cos(k_y) \big] \Gamma_2' + \lambda_x \sin(k_x) \Gamma_3' \nonumber \\
&+ \big[ \gamma_x + \lambda_x \cos(k_x) \big] \Gamma_4' + \lambda_z \sin(k_z) \Gamma_5' + \big[ \gamma_z + \lambda_z \cos(k_z) \big] \Gamma_6' ,
\label{suppl_Ham_octupole}
\end{align}
where $\Gamma_i' = \sigma_3 \otimes \Gamma_i$ for $i=0,1,2,3$ with $\Gamma_i$ being the same set of Gamma matrices appearing in Eq.~\eqref{suppl_Ham_TQI_k}, $\Gamma_4' = \sigma_1 \otimes I_{4 \times 4}$, $\Gamma_5' = \sigma_2 \otimes I_{4 \times 4}$, $\Gamma_6' = i \Gamma_0' \Gamma_1' \Gamma_2' \Gamma_3' \Gamma_4' \Gamma_5'$, and $\gamma_{x,y,z}$ and $\lambda_{x,y,z}$ are intra- and inter-site hopping strengths. When $|\lambda_i| > |\gamma_i|$ for all $i=x,y,z$, the half-filled ground state has topologically nontrivial octupole moment.

\subsubsection{Evaluation of $\langle \hat{U}_3 \rangle$}
\begin{figure}[t!]
\centering\includegraphics[width=0.5\textwidth]{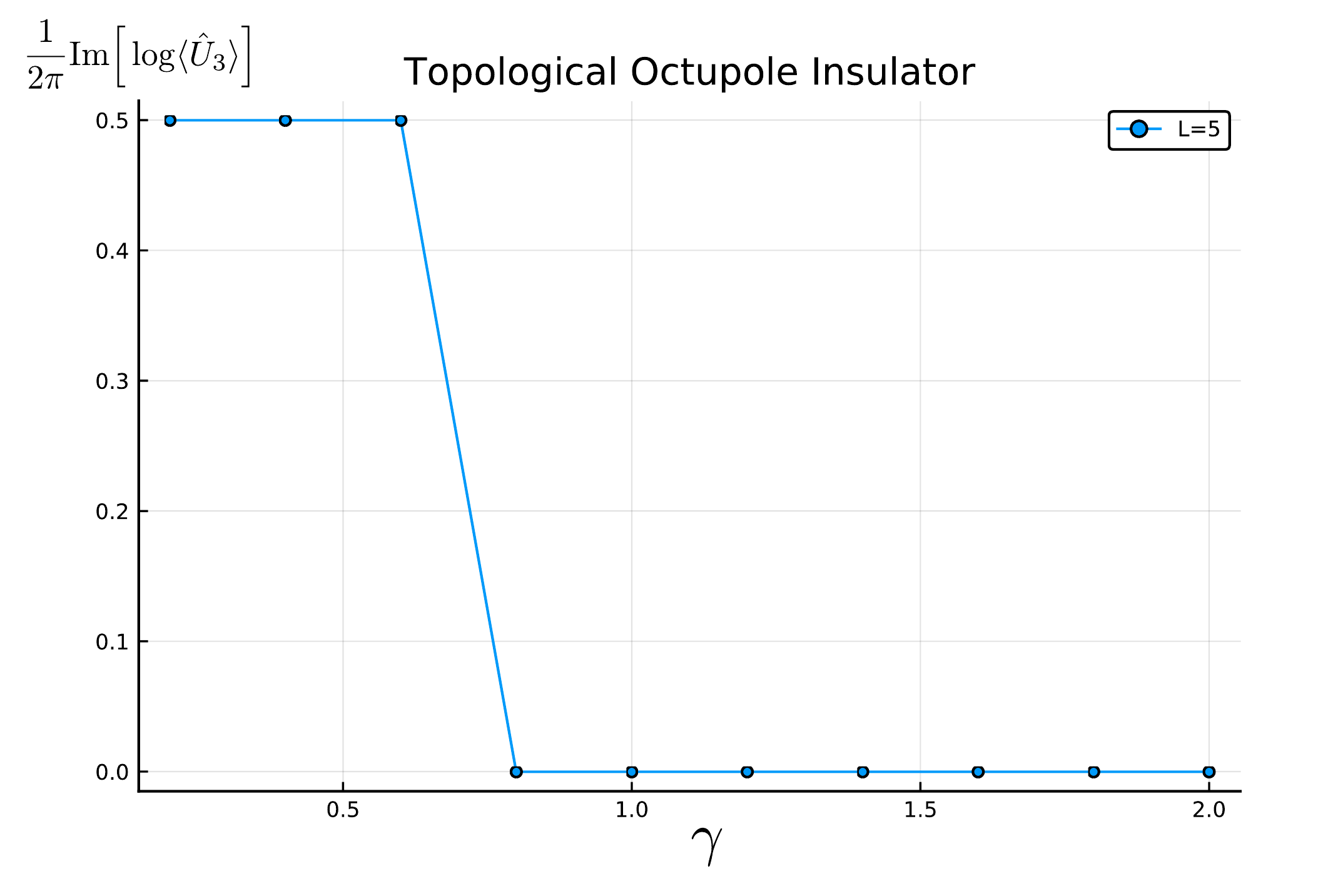}
\caption{The complex phase of $\langle \hat{U}_3 \rangle$ for topological octupole insulator Eq.~\eqref{suppl_Ham_octupole}. We set $\lambda_x = \lambda_y = \lambda_z = 1.0$ and tune $\gamma \equiv \gamma_x = \gamma_y = \gamma_z  \in [0,2]$. When $\gamma < 1.0$ ($\gamma >1.0$), the ground state is in topological (trivial) octupole insulator phase, which is indeed captured by $\langle \hat{U}_3 \rangle$ up to finite-size effect.}
\label{suppl_fig_U3}
\end{figure}

Here we present the numerical evaluation of the expectation value of $\hat{U}_3$ with respect to the half-filled ground state of the octupole insulator Eq.~\eqref{suppl_Ham_octupole}. As done in the quadrupole insulator case, which is summarized in the FIG.~1 in main text, we change the parameter through the phase transition between the trivial insulator and the topological octupole insulator, as shown in FIG.~\ref{suppl_fig_U3}. 

\subsection{Anomalous Topological Quadrupolar Insulator}

A minimal model for an anomalous topological quadrupole insulator is given by Ref. [\onlinecite{PhysRevB.98.201114}].
\begin{align}
H_\textrm{ATQI}(\k) = &\big[ 2t_x\big(1 - \cos (k_x)\big) - \mu \big] \sigma_3 \otimes \sigma_0 \otimes \sigma_0 + V_z \sigma_0 \otimes \sigma_3 \otimes \sigma_0 + \Delta \sigma_1 \otimes \sigma_0 \otimes \sigma_0 \nonumber \\
&+ \alpha \sin (k_x) \sigma_3 \otimes \sigma_2 \otimes \sigma_0 - \big[\beta_1 - \beta_2 \cos(k_y) \big] \sigma_3 \otimes \sigma_1 \otimes \sigma_2 - \beta_2 \sin (k_y) \sigma_3 \otimes \sigma_1 \otimes \sigma_1 ,
\label{suppl_Ham_ATQI}
\end{align}
where $t_x$ is the nearest neighbor hopping strength in $x$-direction, $\mu$ is the chemical potential, $V_z$ is the Zeeman energy, $\Delta$ is the superconducting pairing strength, $\alpha$ and $\beta_1/\beta_2$ are the Rashba spin-orbit coupling strengths in $x$- and $y$-direction. In the main text, we set $(t_x, \mu, \Delta, \alpha, \beta_1, \beta_2) = (1.7, -0.9, 1.6, 3.7, 0.8, 6.2)$ and tune $V_z \in [0.7, 2.7]$. When $V_z > \sqrt{\Delta^2 + \mu'^2}$, the half-filled ground state is topologically nontrivial and when $V_z < \sqrt{\Delta^2 + \mu'^2}$, the half-filled ground state is trivial where $\mu' \approx 0.46$ for our choices of parameters.

\subsection{Edge-Localized Polarization Model}

\begin{figure}[t!]
\centering\includegraphics[width=0.5\textwidth]{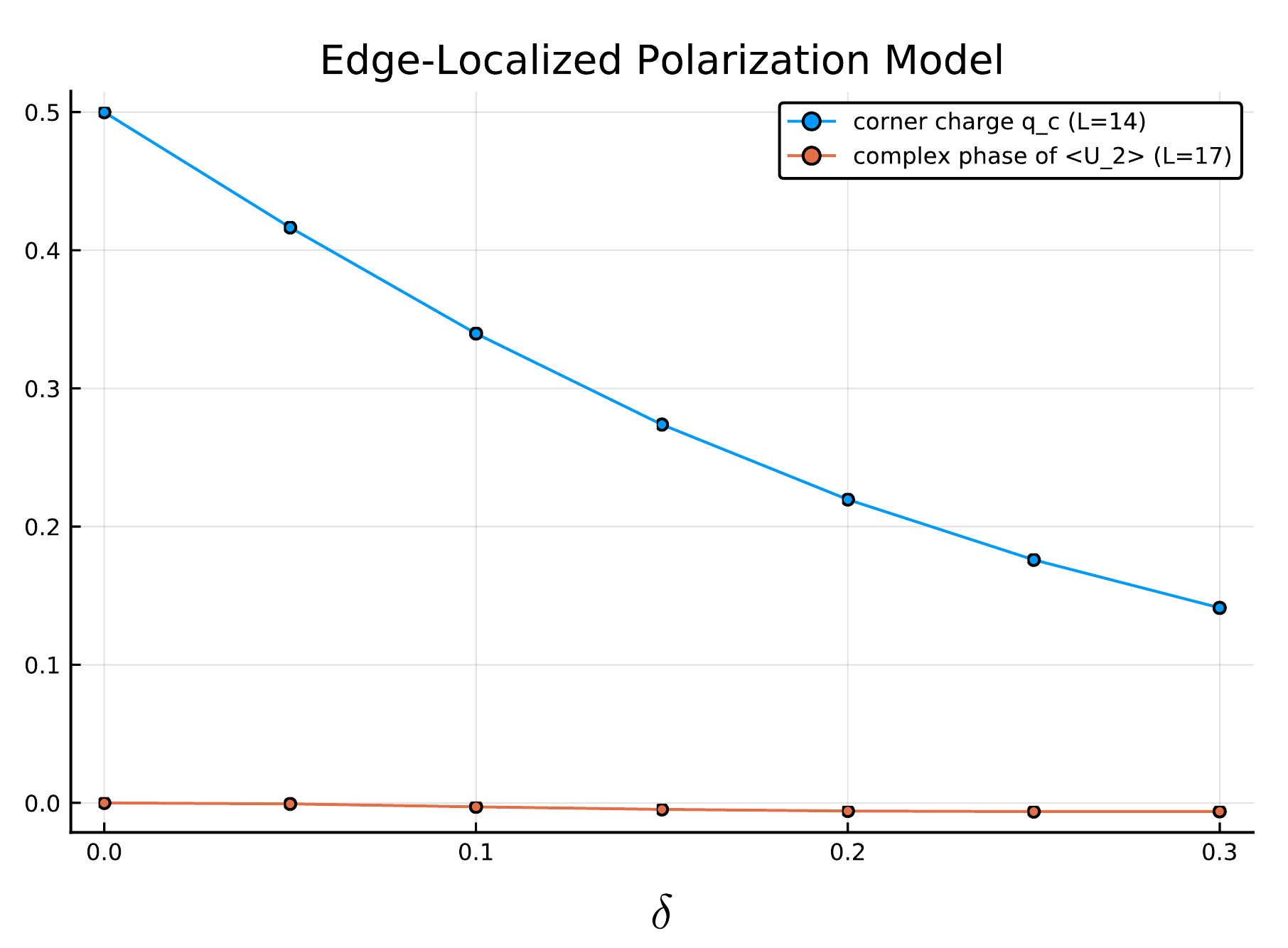}
\caption{Edge-localized polarization model Eq.~\eqref{suppl_Ham_bpol} with $(\gamma, \lambda_x, \lambda_y) = (0.1, 1.0, 0.5)$ and change $\delta \in [0,0.3]$. The corner charge comes solely from the boundary localized polarization and the bulk quadrupole moment vanishes. We indeed see that the complex phase of $\langle \hat{U}_2 \rangle$ is trivial up to finite size effects.}
\label{suppl_fig_bpol}
\end{figure}

One of the key characteristic of the quadrupole insulator is that the following four physical observables are identical: $q_\textrm{c} = |p_x^\textrm{edge}| = |p_y^\textrm{edge}| = Q_{xy}$, where $q_\textrm{c}$ is the corner charge localized at one edge in the case of full open boundary condition, $p_{x(,y)}^\textrm{edge}$ is edge localized polarization along $x$-direction ($y$-direction) in the case of open boundary condition along $x$-direction ($y$-direction) and periodic boundary condition along $y$-direction ($x$-direction), and $Q_{xy}$ is the bulk quadrupole moment. In contrast, there exists model, \textit{edge-localized polarization} model, in which $q_\textrm{c} =  |p_x^\textrm{edge}| + |p_y^\textrm{edge}|$ and $Q_{xy} =0$ instead. A minimal model for this edge-localized polarization model is given by~\cite{PhysRevB.96.245115}
\begin{align}
h(\k) &= \left(
\begin{array}{cc}
\delta \tau_0 & q(\k) \\
q^\dagger(\k) & - \delta \tau_0
\end{array}
\right) , \nonumber \\
q(\k) &= \left(
\begin{array}{cc}
\gamma + \lambda_x e^{i k_x} & \gamma + \lambda_y e^{i k_y} \\
\gamma + \lambda_y e^{-i k_y} & \gamma + \lambda_x e^{-i k_x}
\end{array}
\right) ,
\label{suppl_Ham_bpol}
\end{align}
where $\tau_0$ is the $2 \times 2$ identity matrix and $\gamma$ and $\lambda$ are intra- and inter-layer hopping strengths. 
To see our many-body order parameter $\hat{U}_2$ Eq.~\eqref{suppl_U2} indeed not sensitive to the edge-localized polarization that is not coming from the bulk quadrupole moment, we numerically compute the complex phase of $\langle \hat{U}_2 \rangle$ for the minimal model Eq.~\eqref{suppl_Ham_bpol}. In FIG.~\ref{suppl_fig_bpol}, we set $(\gamma, \lambda_x, \lambda_y) = (0.1, 1.0, 0.5)$ and change $\delta \in [0,0.3]$. As we tune $\delta$, the corner charge changes and equals $|p_x^\textrm{edge}| + |p_y^\textrm{edge}|$, while the bulk quadrupole moment vanishes. As we can be seen in the figure, the complex phase of $\langle \hat{U}_2 \rangle$ is trivial, hence capturing the vanishing bulk quadrupole moment.

\section{Miscellaneous Observations on many-body order parameters}
Here we present several miscellaneous observations on many-body order parameters. This includes the scaling behavior of $|\langle \hat{U}_a \rangle|$ and the dependence on the coordinate re-parametrizations. To be concrete, we concentrate on the polarization $\hat{U}_1$ and quadrupolar moment $\hat{U}_2$. 

\subsection{Scaling of $|\langle \hat{U}_a \rangle |$}


\begin{figure}[t!]
\centering\includegraphics[width=0.7\textwidth]{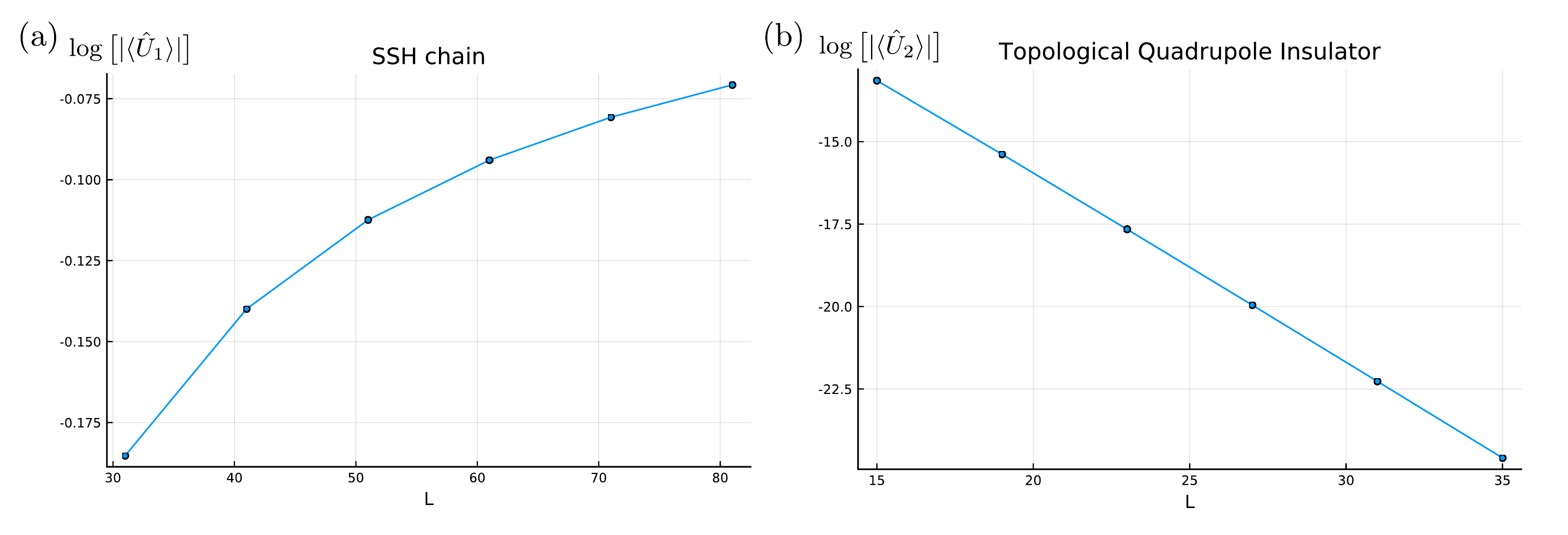}
\caption{Scaling of many-body order parameters $|\langle \hat{U}_{n=1,2} \rangle|$ versus the linear system size $L$ in the case of (a) SSH chain Eq.~\eqref{suppl_Ham_SSH_k} with $(\gamma, \lambda, \delta)=(1.0, 2.0, 0.1)$ and (b) topological quadrupole insulator Eq.~\eqref{suppl_Ham_TQI_k} with $(\gamma_x, \gamma_y, \lambda_x, \lambda_y, \delta)=(1.0, 1.0, 2.0, 2.0, 0.1)$. (a) The modulus of $\langle \hat{U}_1 \rangle$ converges to $1$ as $L \to \infty$, as expected in the case of a gapped insulator. (b) The modulus of $\langle \hat{U}_2 \rangle$ decays exponentially in $L$ even in the case of a gapped insulator.}
\label{suppl_fig_abs_U1_U2}
\end{figure}

\subsubsection{Scaling of $|\langle \hat{U}_1 \rangle|$}

Here we summarize the known scaling forms of the modulus of the expectation value of $\hat{U}_1$, which is introduced by Resta~[\onlinecite{PhysRevLett.80.1800}]. For an one-dimensional system with system size $L$, the Resta's operator is given by
\begin{equation}
\hat{U}_1 = \exp \bigg( \frac{2\pi i}{L} \sum_{x=1}^L x \hat{n}_x \bigg) ,
\label{suppl_U1}
\end{equation}
where $\hat{n}_x = \sum_{\alpha} c_{x, \alpha}^\dagger c_{x, \alpha}$ is the occupation number operator at site $x$ with $\alpha$ being the orbital index and we label the position at site $x$ to $x = 1, 2, \cdots, L$. Eq.~\eqref{suppl_U1} is well-defined in the sense that upon re-labeling a position $x$ by $x+L$, $\hat{U}_1$ remains invariant. Moreover, in the case of a band insulator, the expectation value of $\hat{U}_1$ is related to the Zak phase by (up to an additional minus sign which we address in detail below)
\begin{equation}
\frac{1}{2\pi} \textrm{Im} \big( \log \langle \hat{U}_1 \rangle \big) = \int_\textrm{BZ} dk ~\textrm{tr} \big( \mathcal{A}_k \big) + \mathcal{O} \big( 1/L \big) \mod 1,
\end{equation}
where the expectation value in the LHS is with respect to the many-body ground state, the integral in the RHS is the Zak phase~\cite{PhysRevLett.62.2747} with BZ denotes the one dimensional Brillouin Zone, $\mathcal{A}_k$ is the Berry connection, and $\mathcal{O}(1/L)$ term vanishes as the system size $L \to \infty$. This provides a proof that the phase of the expectation value of $\hat{U}_1$ detects the polarization, as the polarization is equal to the Zak phase in a proper unit.

Although the complex phase of $\langle \hat{U}_1 \rangle$ gives the electric polarization, its modulus also captures invaluable information, e.g., the localization length of the ground state~\cite{PhysRevLett.82.370}. In FIG.~\ref{suppl_fig_abs_U1_U2}, we provide the scaling of (a) $|\langle \hat{U}_1 \rangle|$ for the SSH chain Eq.~\eqref{suppl_Ham_SSH_k} with $(\gamma, \lambda, \delta) = (1.0, 2.0, 0.1)$ versus the linear system size $L$. In this case, the bulk gap is finite. As expected, when the linear system size $L \to \infty$, $|\langle \hat{U}_1 \rangle| \to 1$ in the case of the SSH chain. In general, when the bulk gap is nonzero, the modulus of the expectation value $|\langle U_1 \rangle|$ converges to a positive number as $L \to \infty$. On the other hand, at critical point $|\langle U_1 \rangle|$ vanishes as $L \to \infty$ and it is conjectured to obey the following scaling relation~\cite{PhysRevB.97.165133}:
\begin{equation}
|\langle \hat{U}_1 \rangle| \sim \frac{1}{L^\beta},
\end{equation}
where $\beta >0$ is the exponent characterizing the decay in the thermodynamic limit.

\subsubsection{Scaling of $|\langle \hat{U}_2 \rangle|$}
Here we perform the scaling study of $|\langle \hat{U}_2 \rangle |$ with respect to the length and the Wannier gap. Recall that $\hat{U}_2$ is 
\begin{equation}
\hat{U}_2 = \exp \bigg( \frac{2\pi i}{L_x L_y} \sum_{x=1}^{L_x} \sum_{y=1}^{L_y} x y \hat{n}_{(x,y)} \bigg),
\label{suppl_U2}
\end{equation}
where $\hat{n}_{(x,y)}$ is the occupation number operator at site $(x,y)$ and we label the $x$- ($y$-)position of site $(x,y)$ as $x = 1,2, \cdots, L_x$ ($y = 1,2, \cdots, L_y$). 

First of all, at least within the finite-size calculation up to $L \sim 20$ and within the toy model, we found that, unlike $|\langle \hat{U}_1 \rangle|$, $|\langle \hat{U}_2 \rangle|$ seems to exhibit the exponential decay in $L = L_x = L_y$ as shown in FIG.~\ref{suppl_fig_abs_U1_U2} (b), even for the insulating ground state (here the Wannier bands are also gapped). 
\begin{equation}
|\langle \hat{U}_2 \rangle| \sim e^{- \alpha L},
\end{equation}  
where $\alpha >0$ is an exponent characterizing the exponential decay. \textit{However, to conclude that this scaling behavior is generic for any insulator, we need more detailed and thorough calculations on various models. Keeping this in mind, at this moment we would like to modestly mention that the scaling behaviors of the modulus of $\langle \hat{U}_2 \rangle$ is quite different from those of $\hat{U}_1$.} Despite of exponentially vanishing modulus for the models that we consider, the complex phases of the expectation value can be reliably measured for the typical system sizes that we worked on, e.g., $L \sim 10 - 20$. We also have checked that the Dirac semimetal state, which has a zero energy gap, also shows an exponential decay (with different exponent) in system size $L$. 

\begin{figure}[t]
\centering\includegraphics[width=0.7\textwidth]{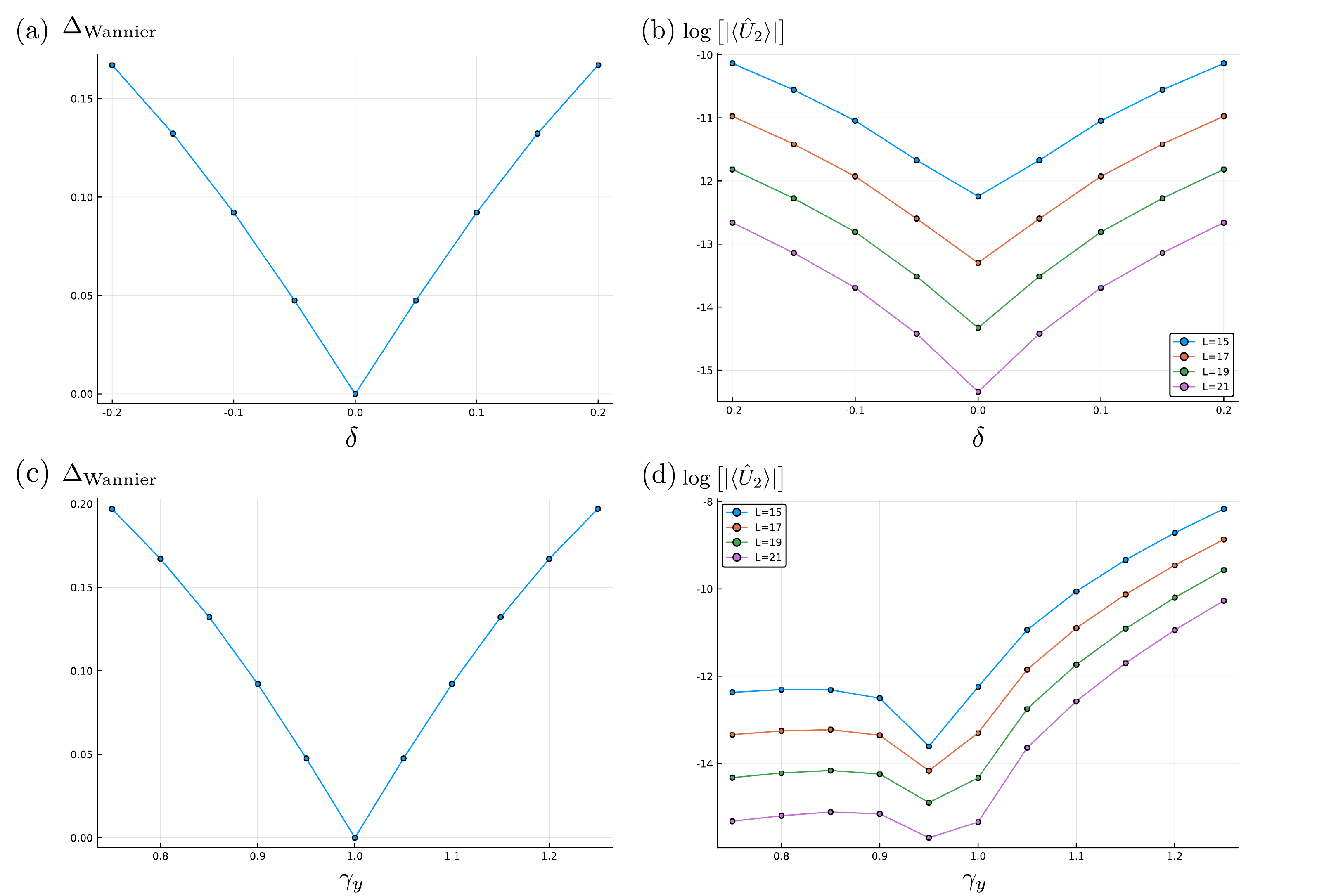}
\caption{Comparison between the Wannier gap and the modulus of $|\langle \hat{U}_2 \rangle|$ for topological quadrupole insulator Eq.~\eqref{suppl_Ham_TQI}. In (a) and (b) we fix $(\gamma_x, \gamma_y, \lambda_y, \lambda_x, \lambda_y) = (0.75, 1.0, 1.0, 1.0)$ and tune $\delta \in [-0.2, 0.2]$. When $\delta = 0$ (a) the Wannier gap closes and (b) $|\langle \hat{U}_2 \rangle|$ is the smallest. In (c) and (d) we fix $(\gamma_x,  \lambda_x, \lambda_y, \delta) = (0.75, 1.0, 1.0, 0.0)$ and tune $\gamma_y \in [0.75, 1.25]$. (c) The Wannier gap associated with the Wannier band $\nu_x(k_y)$ closes at $\gamma_y = 1.0$ and (d) $|\langle \hat{U}_2 \rangle|$ is the smallest around $\gamma_y = 1.0$. In all cases, the Wannier gap is well-defined in the thermodynamic limit while $|\langle \hat{U}_2 \rangle|$ vanishes in the thermodynamic limit.
}
\label{suppl_fig_scaling}
\end{figure}

Second, we perform the comparative studies of the modulus with the Wannier gap. For this we use the model of the topological quadrupole insulator Eq.~\eqref{suppl_Ham_TQI_k} with $\delta=0$. Here the topological-to-trivial quadrupole insulator transition is associated with the Wannier gap closing transition while the bulk gap may not close at the transition point~\cite{PhysRevB.96.245115}. To see how $\langle \hat{U}_2 \rangle$ detects such quantum phase transition, we compute $|\langle \hat{U}_2 \rangle|$ across the phase transition. Due to the quantization by symmetries, $\langle \hat{U}_2 \rangle$ is a real number and changes its sign across the phase transition. So precisely at the transition point, $\langle \hat{U}_2 \rangle$ vanishes up to the correction from finite-size effect. In FIG.~\ref{suppl_fig_scaling}, we present two scenarios of Wannier-gap closing transition, one by tuning $\delta$ and the other by tuning $\lambda_y$. In FIG.~\ref{suppl_fig_scaling} (a) and (b), we fix $(\gamma_x, \gamma_y, \lambda_x, \lambda_y) = (0.75, 1.0, 1.0, 1.0)$ and tune $\delta \in [-0.2, 0.2]$. When $\delta =0$, Wannier gap closes and $|\langle \hat{U}_2 \rangle|$ becomes the smallest. In FIG.~\ref{suppl_fig_scaling} (c) and (d), we fix $(\gamma_x, \lambda_x, \lambda_y, \delta) = (0.75, 1.0, 1.0, 0.0)$ and tune $\gamma_y \in [0.75, 1.25]$. When $\gamma_y < 1$, the half-filled ground state is the topological quadrupole insulator and when $\gamma_y > 1$, the half-filled ground state is trivial. In FIG.~\ref{suppl_fig_scaling} (c), we compute the Wannier gap associated with the Wannier bands $\nu_x(k_y)$. We see that the minimum of $|\langle \hat{U}_2 \rangle|$ occurs around the phase transition point $\gamma_y = 1.0$.

\subsection{Dependence on Coordinate Parameterization}

As outlined in the main text, our many-body operators as well as the original Resta's operator can be generalized by acting only on subsystem and/or change in the boundary conditions. In this subsection, we would like to discuss the dependence on the coordinate parametrization on the many-body operators. To be precise, let us consider the following many-body operator which fully takes into account both the partial action and the coordinate parametrization issue.
\begin{align}
\hat{V}_1 (l, d) = 
\begin{cases}
\exp \Big[ \frac{2\pi i}{l} \sum_x (x-d) (\hat{n}_x - \overline{n}) \Big] \quad \textrm{for } x \in [1, l] \\
1 \qquad \qquad \qquad \qquad \qquad \qquad \quad \textrm{otherwise} 
\end{cases} ,
\label{suppl_V1}
\end{align}
where we consider a finite system with size $L$, $l>0$ is an integer smaller than or equal to $L$, atomic sites are labeled by $x \in \{1,2, \cdots, L\}$, and $\overline{n}$ is the average filling per site. Similarly for quadrupole moment,
\begin{align}
\hat{V}_2 (l, d) = 
\begin{cases}
\exp \Big[ \frac{2\pi i}{l^2} \sum_{\r = (x,y)} (x-d)(y-d) (\hat{n}_{\r} - \overline{n}) \Big] \quad \textrm{for } \r \in [1, l] \times [1, l] \\
1 \qquad \qquad \qquad \qquad \qquad \qquad \qquad \qquad \qquad \textrm{otherwise} 
\end{cases} ,
\label{suppl_V2}
\end{align}
where the whole system size is given by $(L, L)$, $l>0$ is an integer less than or equal to $L$, sites are labeled by $x,y \in \{1, 2, \cdots, L\}$, and $\overline{n}$ denotes the average filling per site. Here, we present the isotropic case, but anisotropic as well as generalization to higher-order moments is immediate. Note that these are almost the same as the formula Eq. (6) and Eq. (7) in the main text, but with an extra phase factors $\propto \exp (2\pi i \bar{n})$, which is the contribution from background positive charges. Here $d$ parametrizes the dependence on the ``origin" of the coordinate systems.

\begin{figure}[h!]
\centering\includegraphics[width=0.7\textwidth]{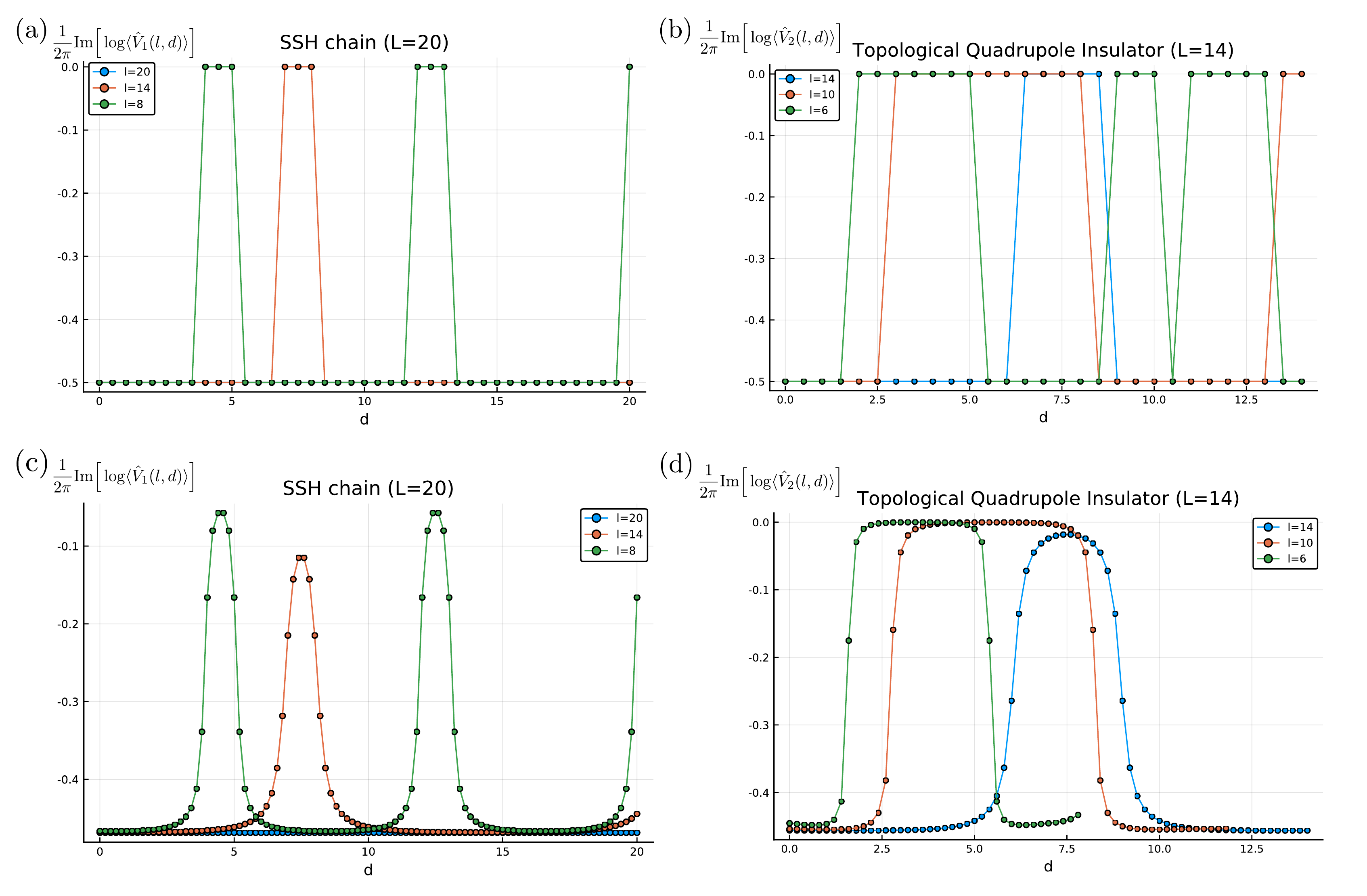}
\caption{The complex phase of $\langle \hat{V}_1(l,d) \rangle$ and $\langle \hat{V}_2(l, d) \rangle$ for various $l$'s as a function of $d$. (a) and (c) correspond to the SSH chain Eq.~\eqref{suppl_Ham_SSH_k} where we set $(\gamma, \lambda) = (1.0, 2.0)$ and (c) $\delta = 0$ and (d) $\delta = 0.1$. (b) and (d) correspond to the topological quadrupole insulator Eq.~\eqref{suppl_Ham_TQI_k} where we set $(\gamma_x, \gamma_y, \lambda_x, \lambda_y) = (1.0, 1.0, 2.0, 2.0)$ and (c) $\delta = 0$ and (d) $\delta = 0.1$. While the ground state of (a) and (b) are topologically nontrivial, for some $l$ and $d$, the complex phase of $\langle \hat{V}_{n} (l,d) \rangle$ becomes trivial. In (c) and (d), as a function of $d$, we see ``lumps''.}
\label{suppl_fig_Vs}
\end{figure}

Now we present numerical evaluation of $\hat{V}_1(l, d)$ Eq.~\eqref{suppl_V1} and $\hat{V}_2(l,d)$ Eq.~\eqref{suppl_V2} in the case of SSH chain Eq.~\eqref{suppl_Ham_SSH_k} and topological quadrupole insulator Eq.~\eqref{suppl_Ham_TQI_k}. Here, $l$ denotes the linear system size within which the operator $V_n(l,d)$ acts nontrivially and $d$ tunes the choice of the origin of our system. In FIG.~\ref{suppl_fig_Vs}, we present the results for $\delta = 0$ and $\delta = 0.1$, where the former has the quantization symmetries and the latter does not. We set $(\gamma, \lambda) = (1.0, 2.0)$ for FIG.~\ref{suppl_fig_Vs} (a) and (c), and set $(\gamma_x, \gamma_y, \lambda_x, \lambda_y) = (1.0, 1.0, 2.0, 2.0)$ for FIG.~\ref{suppl_fig_Vs} (b) and (d). Hence for FIG.~\ref{suppl_fig_Vs} (a) (FIG.~\ref{suppl_fig_Vs} (b)), the ground state has topologically nontrivial polarization (quadrupole moment). 

As expected, when $l=L$, $\langle \hat{V}_1 (l=L, d) \rangle$, which is the original Resta's operator, is independent of $d$ due to the charge conservation in the ground state. However, this is no longer true for $\langle \hat{V}_1 (l<L, d) \rangle$ and $\langle \hat{V}_2 (l, d) \rangle$ for all $l$, which is summarized in FIG.~\ref{suppl_fig_Vs} (a) and (b). As a result, for some choices of $l$ and $d$, the complex phase of $\langle \hat{V}_n (l, d) \rangle$ fails in capturing the ground state topological multipole moments. When $\delta = 0.1$, quantization of $\langle \hat{V}_n (l, d) \rangle$ no longer exists and its complex phase can take arbitrary value, which can be seen in FIG.~\ref{suppl_fig_Vs} (c) and (d).

This behavior can be easily explained by considering the ultra-short correlated states. For the illustrational purpose, we take the topological and trivial ground states of the polarization chain. 
\begin{align}
|GS_{\text{triv}}\rangle = \prod_{n \in \mathbb{Z}} |n \rangle, ~~ |GS_{\text{top}}\rangle = \prod_{n \in \mathbb{Z}} |n + \frac{1}{2} \rangle. 
\end{align}
For these ground states, we see that $\langle GS_{\text{triv}} |\hat{V}_1(l, \frac{l}{2}) |GS_{\text{triv}} \rangle = \langle GS_{\text{top}} |\hat{V}_1(l, \frac{l}{2}) |GS_{\text{top}} \rangle = 1$, signaling that the two states cannot be distinguished if $d = \frac{l}{2}$. Similar discussion can be made for the quadrupolar insulators, too, and this explains the lumps appearing in the numerical data Fig. \ref{suppl_fig_Vs}.

\section{Application of Many-Body Order Parameter $\hat{U}_2$ on Generic $C_4$-Symmetric Insulators}
Recently, a new class of insulators in two dimensions, called $C_n$-symmetric insulators, are introduced~\cite{Cn_insulator} in which a fractional charges can be localized in a corner when the system is subject to full open boundary condition, similar to quadrupole insulators. $C_n$-symmetric insulators can be understood in terms of "filling anomaly" which counts the number of accesive or deficit of electrons between electrons at constant filling and electrons satisfying the charge neutrality~\cite{Cn_insulator}. As a result of a nonzero filling anomaly, fractional charges are localized in all corners in a $C_n$ respecting mannar \textit{only when} the bulk polarization becomes zero. To better understand our many-body operator $\hat{U}_2$ and its relation to nested Wilson loop, let us focus on $C_4$-symmetric cases with zero polarizations in Ref.~\onlinecite{Cn_insulator}. We consider two models -- 1) $h_{1b}^{(4)}+h_{2c}^{(4)}$, a model with average filling $3/8$ and corner charge $1/4$, and 2) $h_{2b}^{(4)}$, a model with $1/2$ filling and corner charge $1/2$, -- where we use the notation introduced in Ref.~\onlinecite{Cn_insulator}. For these states, because they do not posses the gap in the Wannier bands and so do not have the quadrupole order (the nested Wilson loop indices require non-vanishing Wannier gaps), we expect the many-body order parameter, i.e., the phase value $Q_{xy}$ of $\hat{U}_2$, to exhibit unstable continuum values instead of quatized values matching the corner charge. We explicitly confirm these from the calculations below.

\subsection{$h_{1b}^{(4)}+h_{2c}^{(4)}$ model}
$h_{1b}^{(4)}$ is a four-band model which has "corner charge" $e/4$ at $1/4$ filling. $h_{2c}^{(4)}$ is also a four-band model which has 0 "corer charge" at $1/2$ filling. Since both $h_{1b}^{(4)}$ and $h_{2c}^{(4)}$ have polarization ${\bf P} = (\frac{e}{2}, \frac{e}{2})$, we stack them together so that the net polarization becomes 0~\cite{Cn_insulator}. Following Ref.~\onlinecite{Cn_insulator}, we turn on onsite hopping terms between orbitals of $h_{1b}^{(4)}$ and $h_{2c}^{(4)}$. The Bloch Hamiltonian is given by,
\begin{equation}
h_{1/4}^{(4)} ({\bf k}) = \left(
\begin{array}{cc}
h_{1b}^{(4)} ({\bf k}) & \gamma_c \\
\gamma_c^\dagger & h_{2c}^{(4)} ({\bf k}) ,
\end{array} \right)
\label{suppl_h_1_4}
\end{equation}
where 
\begin{equation}
h_{1b}^{(4)} ({\bf k}) = \left( 
\begin{array}{cccc}
 0 & t+e^{ik_x} & 0 & t+e^{ik_y} \\
 t+e^{-ik_x} & 0 & t+e^{ik_y} & 0 \\
 0 & t+e^{-ik_y} & 0 & t+e^{-ik_x} \\ 
 t+e^{-ik_y} & 0 & t+e^{ik_x} & 0 
\end{array} \right) ,
\end{equation}

\begin{equation}
h_{2c}^{(4)} ({\bf k}) = \left( 
\begin{array}{cccc}
 0 & 0& t+1.5e^{ik_x}  & 0 \\
 0 & 0 & 0 & t+1.5e^{ik_y} \\
  t+1.5e^{-ik_x} & 0 & 0 & 0 \\ 
 0 &  t+1.5e^{-ik_y} & 0 &  0
\end{array} \right) ,
\end{equation}
and 
\begin{equation}
\gamma_c = \left( 
\begin{array}{cccc}
 t & t & 0 & 0 \\
 0 & t & t & 0 \\
 0 & 0& t & t \\
 t & 0& 0 & t
\end{array} \right) .
\end{equation}

\begin{figure}[h]
\centering\includegraphics[width=0.8\textwidth]{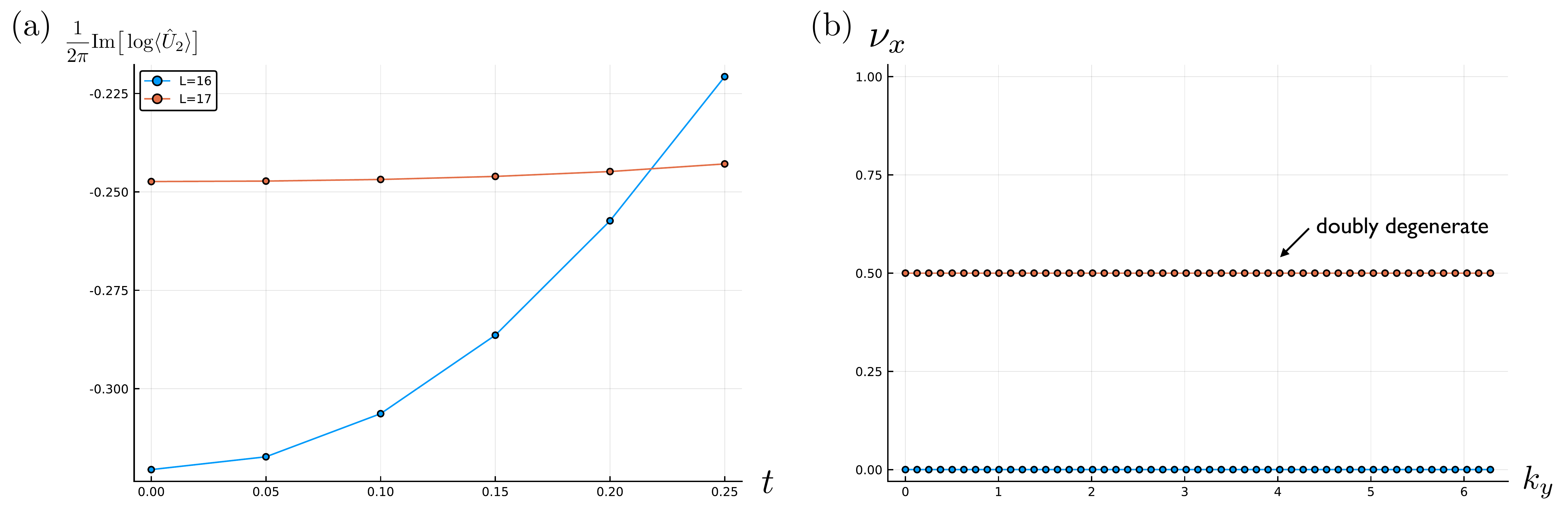}
\caption{(a) $\frac{1}{2\pi} \textrm{Im} \log \langle \hat{U}_2 \rangle$ as a function of hopping parameter $t$ in Eq.~\eqref{suppl_h_1_4} with system sizes $L=16$ and $L=17$. For $L=16$, we take into account additional $(-1)$ factor coming from background charge distribution. Note that both $L=16$ and $L=17$ show dependency in $t$ while the fractional corner charge is independent of $t \in [0, 0.25]$, hence indicating an unstable nature of $C_4$-insulator with respect to $\hat{U}_2$. (b) The Wannier band $\nu_x (k_y)$ of the ground state of Eq.~\eqref{suppl_h_1_4} at $3/8$ filling. We have one doubly degenerate and one non-degenerate flat bands so that in total zero net polarization.}
\label{suppl_fig_h_1_4}
\end{figure}

Under the full open boundary condition, Eq.~\eqref{suppl_h_1_4} at $3/8$ filling have quantized corner charge $3|e|/4$ localized at each corner~\cite{Cn_insulator}. While the corner charge remains quantized and localized when tuning the hopping parameter $t$ from $0$ to $0.25$, the exepctation value of $\hat{U}_2$ is not, as shown in FIG.~\ref{suppl_fig_h_1_4} (a) where strong even-odd effect is also shown. The Wannier band is shown in FIG.~\ref{suppl_fig_h_1_4} (b), where two degerate and one non-degerate flat bands are shown. This indicates that  Eq.~\eqref{suppl_h_1_4} at $3/8$ filling does not have a stable quadrupole moment in the bulk, which is well reflected in our many-body operator $\hat{U}_2$. Note also that the Wannier gap closes, so that the nested Wilson loop indices are not well defined~\cite{PhysRevB.96.245115}, which is consistent with the unstable behavior of $\hat{U}_2$ with respect to the ground state of Eq.~\eqref{suppl_h_1_4} at $3/8$ filling.

\subsection{$h_{2b}^{(4)}$ model}
\begin{figure}[t!]
\centering\includegraphics[width=0.8\textwidth]{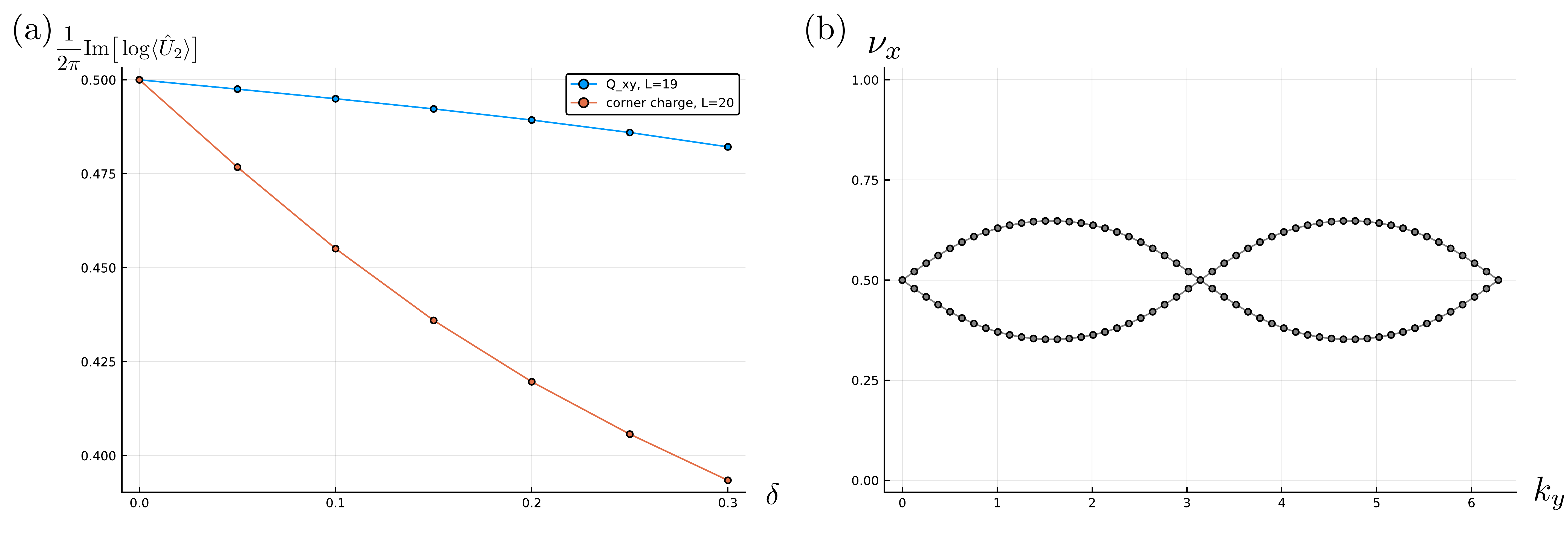}
\caption{(a) $\frac{1}{2\pi} \textrm{Im} \log \langle \hat{U}_2 \rangle$ as a function of onsite $C_4$-breaking parameter $\delta$ in Eq.~\eqref{suppl_h_1_2} with $t=0.3$. Because of finite $\delta$, corner charge changes smoothly from a quantized value $0.5$. Note that there is a mismatch between the corner charge and $\frac{1}{2\pi} \textrm{Im} \log \langle \hat{U}_2 \rangle$, showing an unstable nature of $C_4$-symmetric insulator with respect to $\hat{U}_2$. (b) The Wannier band $\nu_x(k_y)$ of the ground stae of Eq.~\eqref{suppl_h_1_2} with $t=0.3$ and $\delta=0.3$ and at $1/2$ filling. The Wannier band becomes gapless at $k_y=0, \pi$.}
\label{suppl_fig_h_1_2}
\end{figure}

$h_{2b}^{(4)}$ is a four-band model which has corner chage $e/2$ and vanishing polarization ${\bf P} = 0$ at $1/2$ filling. The Bloch Hamiltonian is given by
\begin{equation}
h_{1/2}^{(4)} ({\bf k}) \equiv h_{2b}^{(4)} ({\bf k}) = \left(
\begin{array}{cccc}
\delta & t & e^{i(k_x + k_y)} & t \\
t & -\delta & t & e^{i(k_y-k_x)} \\
e^{-i(k_x + k_y)} & t & \delta& t \\
t & e^{i(k_x-k_y)} & t & -\delta
\end{array}
\right), 
\label{suppl_h_1_2}
\end{equation}
where $t$ is the intra-site hopping parameter and $\delta$ is the strength of an onsite $C_4$-symmetry breaking term. When $0<t<1$ and $\delta=0$, the ground state at half-filling is a $C_4$-symmetric insulator having zero polarization as well as $1/2$ corner charge localized at each corner. Moreove, $\frac{1}{2\pi} \textrm{Im} \log \langle \hat{U}_2 \rangle$ equals $0.5$ ($0$) when $L$ is odd (even), as in the case of topological quadrupole insulators. However, upon introucing symmetry breaking term, i.e., $\delta \ne 0$, there exists a mismatch between $\frac{1}{2\pi} \textrm{Im} \log \langle \hat{U}_2 \rangle$ and the corner charge as shown in FIG.~\ref{suppl_fig_h_1_2} (a). The Wannier band is shown in FIG.~\ref{suppl_fig_h_1_2} (b), where we see that the Wannier band is gapless at $k_y=0$ and $\pi$. Since the Wannier bands are gapless the nested Wilson loop indices are not well defined~\cite{PhysRevB.96.245115}, which is consistent with the unstable behavior of $\hat{U}_2$ with respect to the ground state of  Eq.~\eqref{suppl_h_1_2} at $1/2$ filling.

\bibliography{Multipole-Resta-Formula}